%% file: PASPmanuscript_section_edit.tex
\documentclass[twocolumn]{aastex701}
\usepackage{etoolbox}
\usepackage{fontawesome5}

\newcommand{\Oiii}{\textrm{[\ion{O}{3}]}}
\newcommand{\Oiirec}{\textrm{\ion{O}{2}}}
\newcommand{\Oii}{\textrm{[\ion{O}{2}]}}

\newcommand{\Nii}{\textrm{[\ion{N}{2}]}}

\newcommand{\Hii}{\ion{H}{2}}

\newcommand{\Op}{\ensuremath{\mathrm{O}^+}}
\newcommand{\Opp}{\ensuremath{\mathrm{O}^{++}}}
\newcommand{\Hep}{\ensuremath{\mathrm{He}^+}}

\newcommand{\Hp}{\ensuremath{\mathrm{H}^+}}
\newcommand{\Np}{\ensuremath{\mathrm{N}^+}}

\newcommand*\diff{\mathop{}\!\mathrm{d}}

\newcommand{\aterm}[5]{%
  \ensuremath{#1}\,${}^{#2}\mathrm{#3}%
  \ifx#5o^{\circ}\fi%
  _{#4}$%
}

\defcitealias{campbell86}{C86}

\begin{document}

\title{Revisiting Evidence of Temperature Fluctuations in H II Regions. I. Interionic temperature differences}

\author[0000-0003-4520-5395]{Yuguang Chen}
\affiliation{Department of Physics, The Chinese University of Hong Kong, Shatin, N.T., Hong Kong SAR, China}
\email{yuguangchen@cuhk.edu.hk}

\begin{abstract}

The oxygen abundance of \Hii\ regions is a fundamental tracer of chemical enrichment and gas cycling and is now routinely measured across a broad range of cosmic time. A persistent complication is the abundance discrepancy factor (ADF): the \Opp/\Hp\ abundance inferred from optical recombination lines (RLs) is systematically higher than that inferred from collisionally excited lines (CELs). Small-scale temperature fluctuations within the \Opp-emitting gas are among the most widely discussed explanations for this discrepancy. However, the recent identification of a substantial continuum-pumping contribution to the \Oiirec\ spectrum of M42 motivates a renewed assessment of which observations uniquely favor a temperature-based interpretation and which are also compatible with other mechanisms. In this first paper of a series, I examine the relation between the inferred temperature-fluctuation parameter, $t^2$, and interionic temperature differences, focusing on $\Delta T_e=T_e(\Oiii)-T_e(\Nii)$. Photoionization calculations show that ionization stratification naturally produces systematic temperature differences between the low- and high-ionization zones. Meanwhile, the conversion from an ADF to $t^2$ is strongly temperature dependent because CEL emissivities are nonlinear functions of electron temperature. I show that these two effects can account for much of the observed $t^2$--$\Delta T_e$ relation without requiring the interionic temperature difference itself to trace fluctuations within the \Opp\ zone. This result does not exclude temperature fluctuations in \Hii\ regions. Instead, it indicates that interionic temperature differences should be interpreted with care and, on their own, provide limited discrimination among proposed origins of the ADF.

\end{abstract}


\section{Introduction} \label{sec:intro}

\Opp\ abundances in ionized nebulae are commonly measured using either collisionally excited lines (CELs) or optical recombination lines (RLs). These methods frequently give discrepant results: the abundance inferred from the \Oiirec\ V1 RL multiplet is typically higher than that inferred from the \Oiii\ CEL lines. The logarithmic difference is conventionally described by the abundance discrepancy factor,
\begin{equation}
{\rm ADF}(\Opp)=
\log(\Opp/\Hp)_{\rm RL}-\log(\Opp/\Hp)_{\rm CEL}.
\label{eq:adf}
\end{equation}
For star-forming \Hii\ regions, a representative value is about 0.2--0.3 dex \citep[e.g.,][]{peimbert93, esteban09, esteban14}. 

Temperature fluctuations are among the most commonly invoked explanations for the ADF \citep{peimbert67}. The physical basis of this interpretation is the strong, nonlinear sensitivity of CEL emissivities to electron temperature. A useful approximation for a CEL at wavelength $\lambda$ is
\begin{equation}
\varepsilon_{\rm CEL}(T_e)=C_\lambda T_e^{-A_\lambda}
\exp\left(-\frac{B_\lambda}{T_e}\right),
\label{eq:cel_emissivity}
\end{equation}
where $T_e$ is electron temperature, and $A_\lambda$, $B_\lambda$, and $C_\lambda$ depend on the transition and adopted atomic data \citep{osterbrock06}. Higher-temperature gas can therefore contribute disproportionately to an auroral line. In a nebula with an unresolved temperature distribution, a temperature-sensitive ratio such as \Oiii\ $\lambda4363/\lambda5007$ is biased toward hotter material. If the inferred temperature is then applied to \Oiii\ $\lambda5007/{\rm H}\beta$, the adopted \Oiii\ emissivity can be too large, causing the CEL-based \Opp/\Hp\ abundance to be underestimated.

\begin{figure*}[ht!]
\centering
\includegraphics[width=0.9\textwidth]{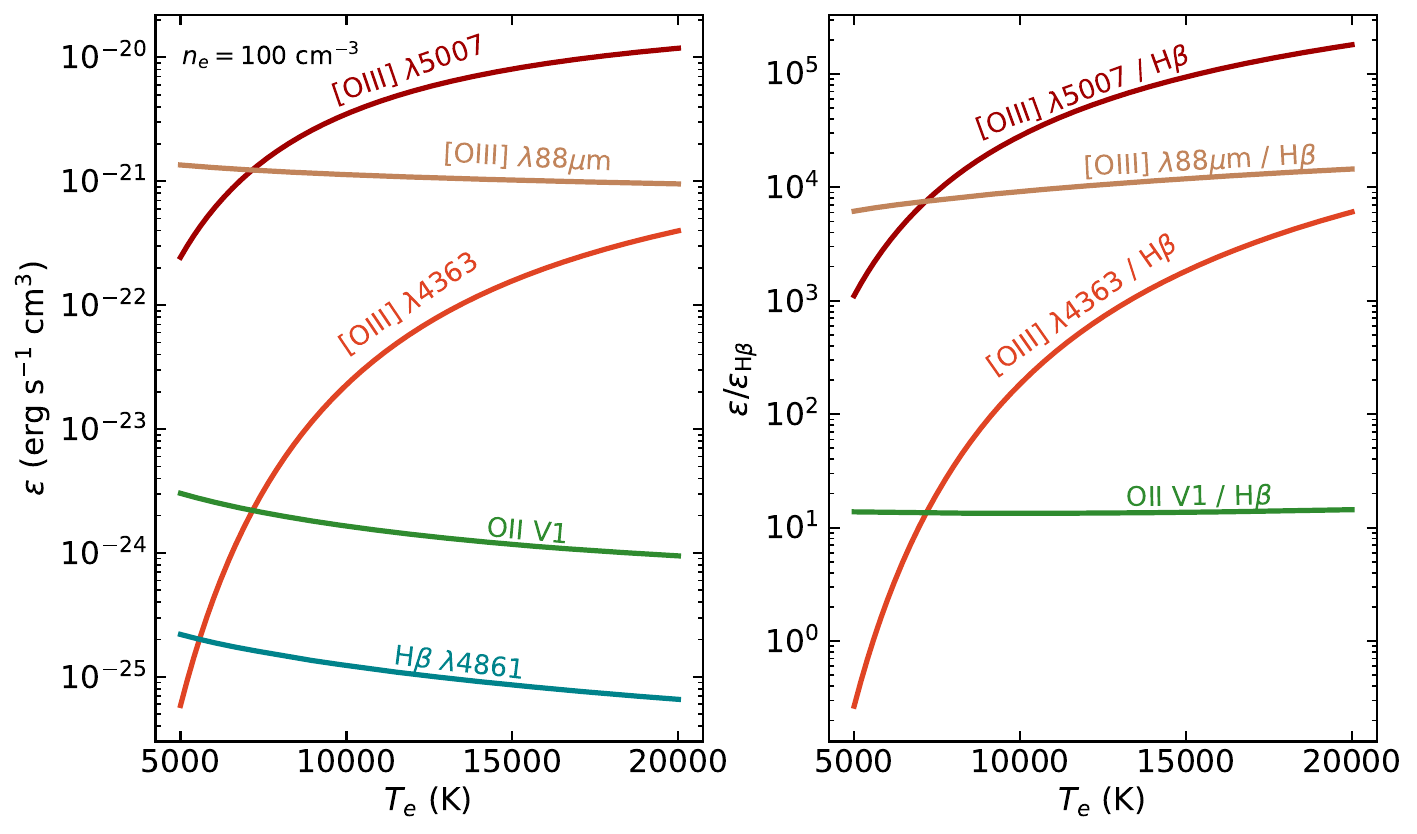}
\caption{\textit{Left:} Reduced emissivities, $\varepsilon$, as functions of electron temperature for selected \Opp\ lines and H$\beta$. \textit{Right:} Corresponding \Opp-to-H$\beta$ emissivity ratios. Both panels assume $n_e=100~{\rm cm}^{-3}$ and Case B recombination and were calculated with \textsc{PyNeb} using atomic data from \citet{storey95,froese-fischer04,storey00,storey14,storey17}. The distinct temperature dependences of the CELs, the \Oiirec\ V1 multiplet, and H$\beta$ form the basis of temperature-based explanations of the ADF.}
\label{fig:emisivity}
\end{figure*}

The temperature dependence of an RL is much shallower and may be approximated as
\begin{equation}
\varepsilon_{\rm RL}(T_e)=E_\lambda T_e^{-D_\lambda},
\label{eq:rl_emissivity}
\end{equation}
where $D_\lambda$ and $E_\lambda$ again depend on the transition and atomic data. RL emission weights cooler gas somewhat more strongly than CEL emission. More importantly, because both \Oiirec\ and H$\beta$ are recombination lines, their ratio is almost a constant as a function of temperature. If the \Opp/\Hp\ abundance is spatially uniform and both features arise purely from recombination, the RL-based abundance is therefore much less sensitive to an unresolved temperature distribution than the CEL-based abundance.

In the classical formalism \citep{peimbert67}, the thermal structure of the volume ($V$) occupied by an ion $X^{i+}$ is characterized by an ion-weighted mean temperature,
\begin{equation}
T_0(X^{i+})=
\frac{\int T_e n_e n(X^{i+}),\diff V}
{\int n_e n(X^{i+}),\diff V},
\label{eq:t0}
\end{equation}
where $n_e$ and $n(X^{i+})$ are the electron and ion number densities, and by a dimensionless mean-square temperature fluctuation,
\begin{equation}
t^2(X^{i+})=
\frac{\int [T_e-T_0(X^{i+})]^2 n_e n(X^{i+}),\diff V}
{T_0(X^{i+})^2\int n_e n(X^{i+}),\diff V}.
\label{eq:t2}
\end{equation}
Thus, for a Gaussian temperature distribution, the standard deviation is $\sigma_T=T_0\sqrt{t^2}$. Values inferred for \Hii\ regions commonly lie in the approximate range $t^2\sim0.01$--0.1. Because different ions generally occupy different volumes, both $T_0$ and $t^2$ must, in principle, be specified for the ionic phase under consideration.

A related explanation involving strong temperature contrasts, invoked particularly for planetary nebulae (PNe), attributes the ADF to cool, hydrogen-deficient, metal-rich inclusions \citep[e.g.,][]{liu00, liu06IAUS, tsamis03, stasinska07}. Such inclusions can contribute little to optical CELs while contributing substantially to heavy-element RLs, producing a large ADF even when a single Gaussian temperature distribution poorly describes the gas and the volume-weighted $t^2$ defined by Equation~\ref{eq:t2} is not especially large. The fluctuation and inclusion pictures are therefore physically distinct, but both exploit the different temperature sensitivities of CEL and RL emission. Under either interpretation, the RL abundance is generally taken to be closer to the true abundance than the uncorrected CEL result.

Temperature structure is not the only process capable of altering the \Opp\ spectrum. Using integral-field spectroscopy of M42, \citet{chen26orion} measured $\sim20$ \Oiirec\ lines and found their relative intensities to be consistent with continuum pumping (fluorescence). In this mechanism, extreme-ultraviolet photons near $400$~\AA\ are absorbed by \Op\ and subsequently re-emitted through \Oiirec\ cascades \citep{grandi76, escalante12}. In the RL analysis, this contribution is incorrectly interpreted as recombination emission. After accounting for continuum pumping, the inferred \Oiirec\ abundance in M42 is significantly reduced, and the measured ADF is consistent with zero within the reported uncertainties.

The M42 result demonstrates that an ADF need not arise exclusively from temperature structure. It neither implies that continuum pumping must dominate in every \Hii\ region nor, by itself, excludes temperature fluctuations. It does, however, motivate a careful reassessment of evidence previously interpreted as favoring temperature fluctuations or chemically distinct cold gas. In particular, it is useful to distinguish observations that compare different thermal weightings within approximately the same ionic volume from those that compare ions occupying different regions of the nebula.

This work focuses on the latter case: the relation between an inferred $t^2$ and an interionic temperature difference, especially
\begin{equation}
\Delta T_e=T_e(\Oiii)-T_e(\Nii).
\label{eq:delta-te}
\end{equation}
Correlations between the ADF and temperature differences have long been studied in planetary nebulae and interpreted in the context of hydrogen-deficient inclusions. More recently, a correlation between $t^2(\Opp)$ and $T_e(\Oiii)-T_e(\Nii)$ has been interpreted as evidence for inhomogeneous temperature distributions in \Hii\ regions \citep{mendez-delgado23nature}. The central question considered here is narrower: does this correlation require the interionic temperature difference to trace fluctuations within the \Opp\ gas, or can it instead arise from ordinary ionization stratification combined with the manner in which $t^2$ is inferred?

This paper is organized as follows. \S2 reviews the historical development of comparisons between the ADF and interionic temperature differences. \S3 examines the temperature structure predicted by photoionization models, the empirical $T_2$--$T_3$ relation, and the nonlinear conversion between the ADF and $t^2$, and then constructs a simple null model for the observed correlation. \S4 considers the physical interpretation and observational tests that could better distinguish structural temperature gradients from fluctuations within a single ionic phase. \S5 summarizes the conclusions.

The quantitative analysis is restricted to star-forming \Hii\ regions. PNe can instead exhibit much larger ADFs, and their discrepancies have been associated with properties of the central stars, including binarity \citep[e.g.][]{wesson18}. The distributions of ADF, temperature difference, and ionizing-source properties therefore suggest that the dominant mechanism may not be the same in planetary nebulae and \Hii\ regions. Planetary-nebula results remain important to the historical discussion but are not included in the subsequent \Hii-region correlation analysis. All code used in this work to generate the figures is available on \href{https://github.com/yuguangchen1/temperature-fluctuations-structure}{GitHub~\faGithub}.

\section{The ADF and Temperature Differences: Historical Context}
\label{sec:historical-context}

Temperature differences have been used to investigate unresolved thermal structure since the introduction of the temperature-fluctuation formalism. \citet{peimbert67} compared electron temperatures derived from temperature-sensitive CEL ratios, including \Oiii\ $\lambda4363/\lambda5007$, with those derived from the hydrogen Balmer discontinuity in three regions of the Orion Nebula. The CEL temperatures were systematically higher. Because \Opp\ and \Hp\ occupy broadly overlapping volumes in an \Hii\ region, this difference can be interpreted as arising from a nonuniform temperature distribution: the \Oiii\ auroral-to-nebular ratio preferentially weights hotter material, whereas the Balmer discontinuity is weighted toward cooler recombining gas. The same framework offered an explanation for the discrepancy between \Opp/\Hp\ abundances inferred from \Oiii\ CELs and \Oiirec\ RLs, which had already been recognized by \citet{wyse42}.

The Balmer jump and \Oiii\ line ratios sample overlapping gas and respond differently to its temperature. Hydrogen recombination emission comes from throughout the ionized nebula, including the \Opp\ zone that produces \Oiii\ emission. Their temperature difference can therefore reveal thermal variations within gas that contributes to both diagnostics. This overlap matters when interpreting later correlations with the ADF, because temperatures measured from ions in different zones can differ simply because they sample different parts of the nebula.

Subsequent studies examined these relations systematically, particularly in planetary nebulae. \citet{liu01,tsamis04,liuy04,wesson05} reported strong correlations between the ADF and $T_e(\Oiii)-T_e({\rm BJ})$. These comparisons were later extended to other recombination-weighted diagnostics, including temperatures inferred from \ion{He}{1} and \Oiirec\ emission \citep{liuy04,mendez-delgado25}. Although these diagnostics all contrast the strong temperature sensitivity of a CEL ratio with a recombination-weighted temperature, they do not weight the nebula identically. Their quantitative relations with the ADF therefore do not necessarily have the same normalization or slope.

One instructive example is the commonly observed ordering $T_e(\textrm{\ion{He}{1}})<T_e({\rm BJ})<T_e(\Oiii)$ in planetary nebulae. Correspondingly, the ADF correlates not only with $T_e(\Oiii)-T_e({\rm BJ})$, but also with the generally larger temperature difference $T_e(\Oiii)-T_e(\textrm{\ion{He}{1}})$. \citet{zhangy05} showed that $T_e(\textrm{\ion{He}{1}})$ is systematically lower than the hydrogen recombination temperature and argued that this separation is difficult to reproduce with chemically homogeneous models containing only temperature fluctuations. In a two-component picture, by contrast, cool, hydrogen-deficient, metal-rich inclusions can contribute disproportionately to \ion{He}{1} and heavy-element RL emission while contributing less to the hydrogen recombination spectrum. The observed separation between the \ion{He}{1} and hydrogen recombination temperatures therefore provided additional support for chemically inhomogeneous, two-abundance models of planetary nebulae.

\citet{garcia-rojas07} extended the comparison between the ADF and $T_e(\Oiii)-T_e({\rm BJ})$ to star-forming \Hii\ regions. The relation in their \Hii-region sample was substantially less pronounced ($0.061\pm0.021$ dex per 1000 K, $r=0.575$) than the planetary-nebula trend ($0.166\pm0.020$ dex per 1000 K, $r=0.89$). They therefore cautioned that the abundance discrepancy may not have the same dominant physical origin in \Hii\ regions and planetary nebulae.

More recently, \citet{mendez-delgado23nature} extended the analysis in \Hii\ regions from a CEL--recombination comparison to a CEL--CEL comparison. They found a strong correlation between the inferred $t^2(\Opp)$ and $T_e(\Oiii)-T_e(\Nii)$ and interpreted it as evidence that an inhomogeneous temperature distribution---whether represented by classical temperature fluctuations or by a cooler, metal-rich component---contributes to the ADF. 

The earlier Balmer-jump comparisons used diagnostics that sample broadly overlapping volumes. In the CEL--CEL comparison, however, \Oiii\ and \Nii\ emission arise predominantly in high- and low-ionization zones, respectively, so ionization stratification can produce a temperature difference even when the temperature varies little within the \Opp\ zone. The inferred $t^2(\Opp)$ also depends on $T_e(\Oiii)$ through the temperature-dependent conversion from the ADF. In Section~\ref{sec:structural-origin}, I examine how these two effects can generate a correlation between inferred $t^2(\Opp)$ and the interionic temperature difference.

\section{A Structural Origin of the \ensuremath{t^2}--\ensuremath{\Delta T_e} Relation}
\label{sec:structural-origin}

\subsection{Ionization Stratification and the $T_2$--$T_3$ Relation}
\label{sec:t2-t3}

An ionization-bounded nebula is not isothermal. The incident spectrum is progressively filtered with depth, while the balance among photoelectric heating, metal-line cooling, and recombination cooling varies with ionization state. Because different ions dominate at different depths, these large-scale thermal gradients produce systematic differences among temperatures inferred from their emission lines.

I illustrate this effect with a grid of star-forming \Hii-region models calculated using \textsc{Cloudy} v22.01 \citep{ferland13} and the SF prescription of \citet{ji20}. The models assume plane-parallel geometry, constant gas pressure, and an initial hydrogen density of $\log(n_{\rm H}/{\rm cm}^{-3})=1.15$. The grid points are $[\mathrm{O/H}]=-1.3,-0.7,-0.4,0.0,+0.3$ and $\log U=-3.0$ to $-2.0$ in increments of 0.5 dex. I adopt the solar abundance pattern of \citet{grevesse10}, with $12+\log({\rm O/H})_\odot=8.69$. Helium, nitrogen, and carbon follow the metallicity-dependent prescriptions of \citet{ji20}, including secondary nitrogen enrichment and fixed $\log({\rm C/N})=1.03$. The models include standard interstellar grains, metallicity-scaled depletions, and the local cosmic-ray background.

The ionizing spectra were generated with \textsc{Starburst99} v7.0.1 \citep{leitherer99} for continuous star formation at $1\,M_\odot,{\rm yr}^{-1}$ and an age of 4 Myr, using a Kroupa initial mass function \citep{kroupa01} from 0.1 to $100,M_\odot$. Stellar and gas metallicities are matched. Each calculation is iterated to convergence and stopped when $T_e$ falls below 1000 K, yielding a radiation-bounded model. Line emission is then integrated layer by layer using the local ionic densities and recalculated with \textsc{PyNeb} emissivities. I use \textsc{PyNeb} rather than the integrated \textsc{Cloudy} line fluxes to avoid systematic differences between the emissivity calculations of the two codes. In this way, any differences in the recovered $T_e$ arise from the modeled temperature structure rather than from mismatched atomic emissivities.

\begin{figure*}[ht!]
\centering
\includegraphics[height=8cm]{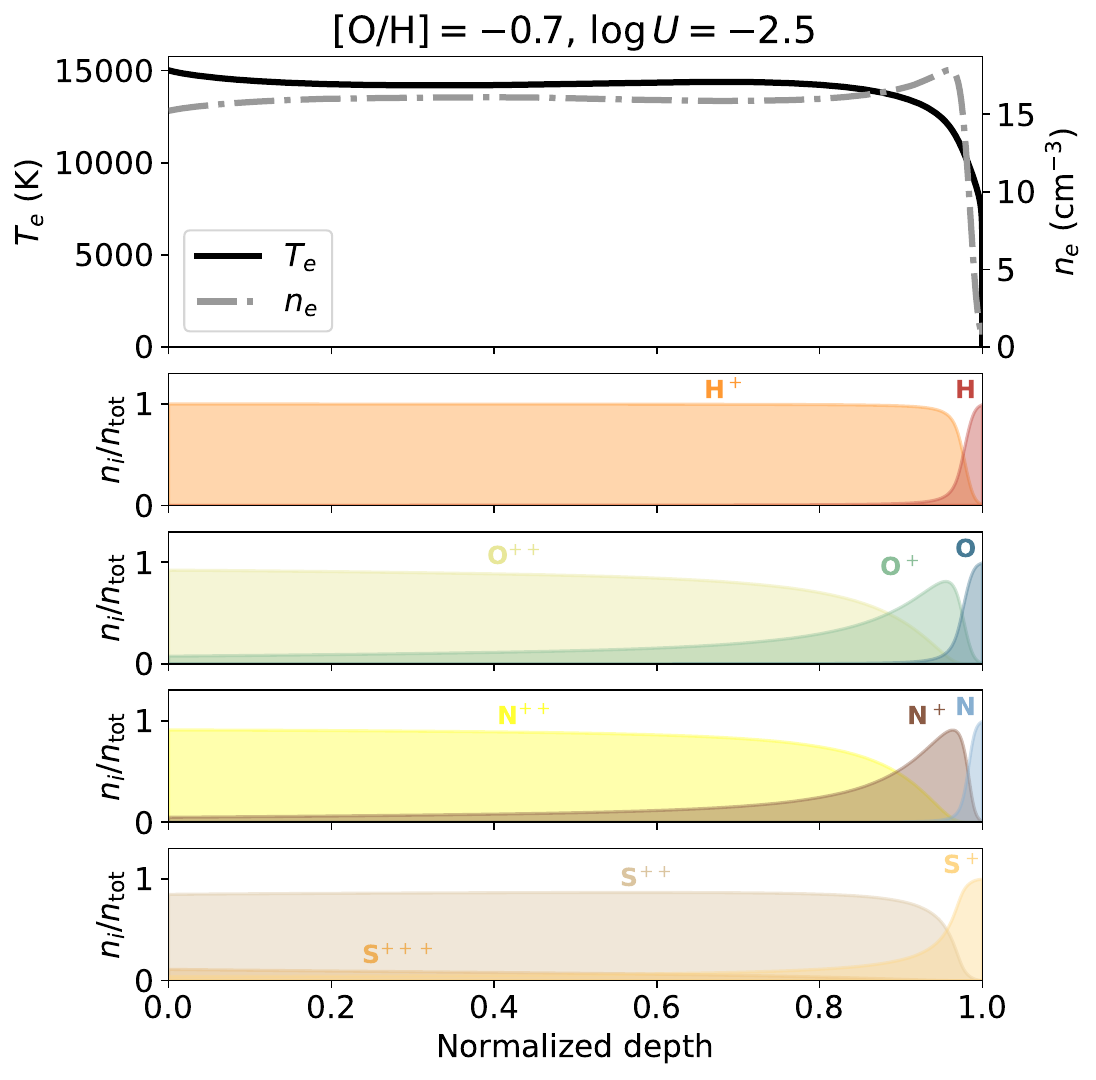}
\includegraphics[height=8cm]{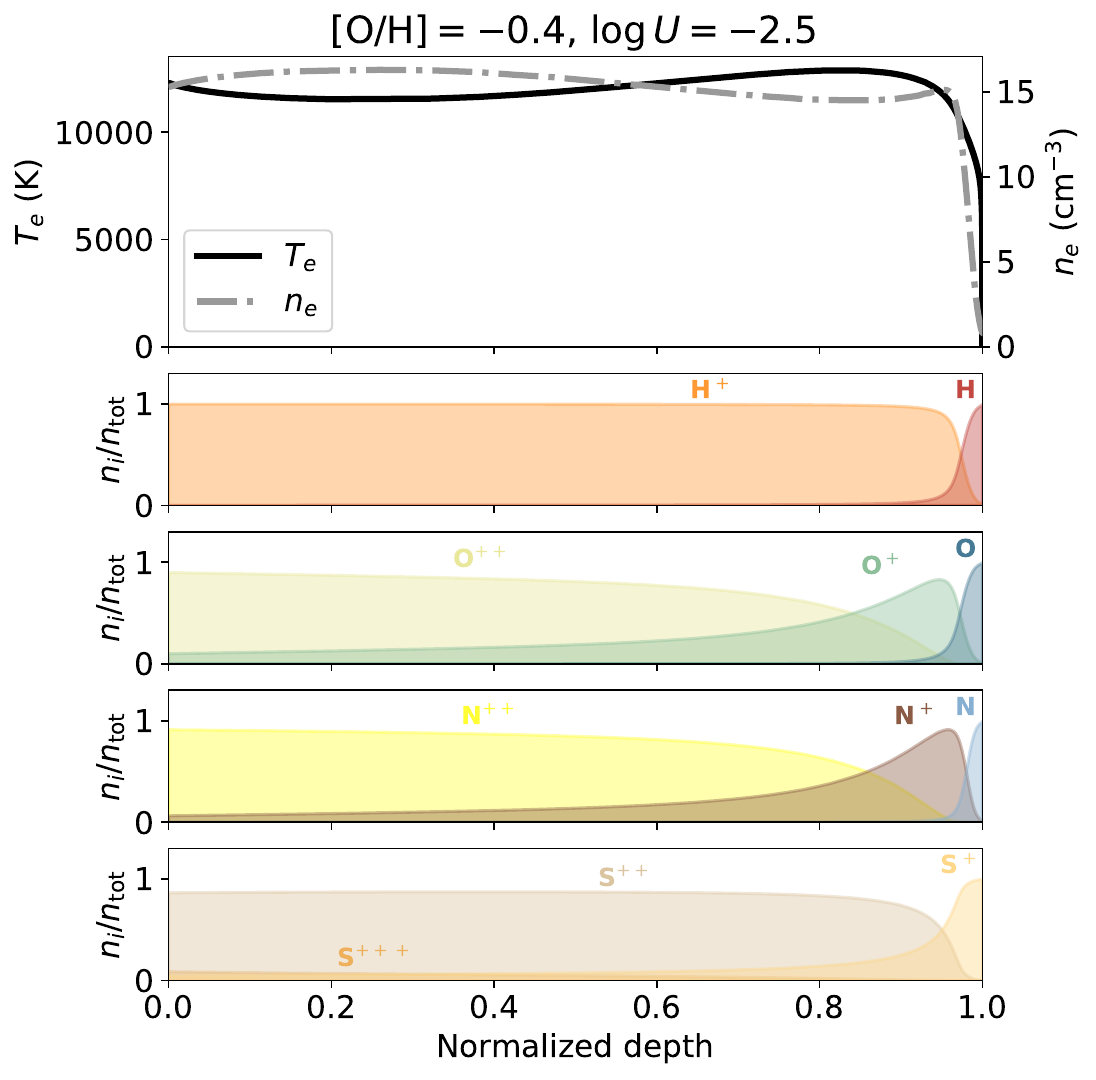}\\
\includegraphics[height=8cm]{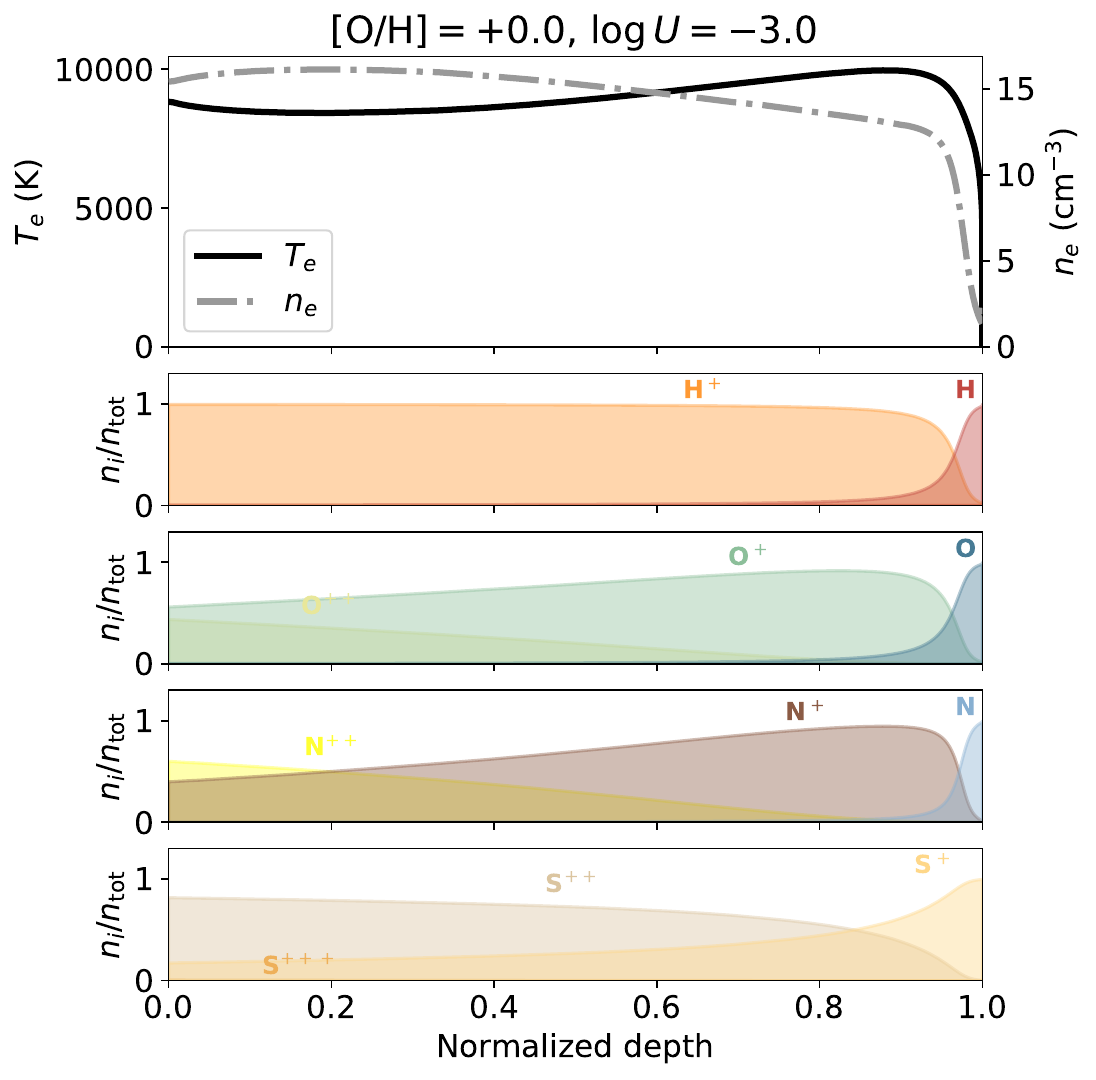}
\includegraphics[height=8cm]{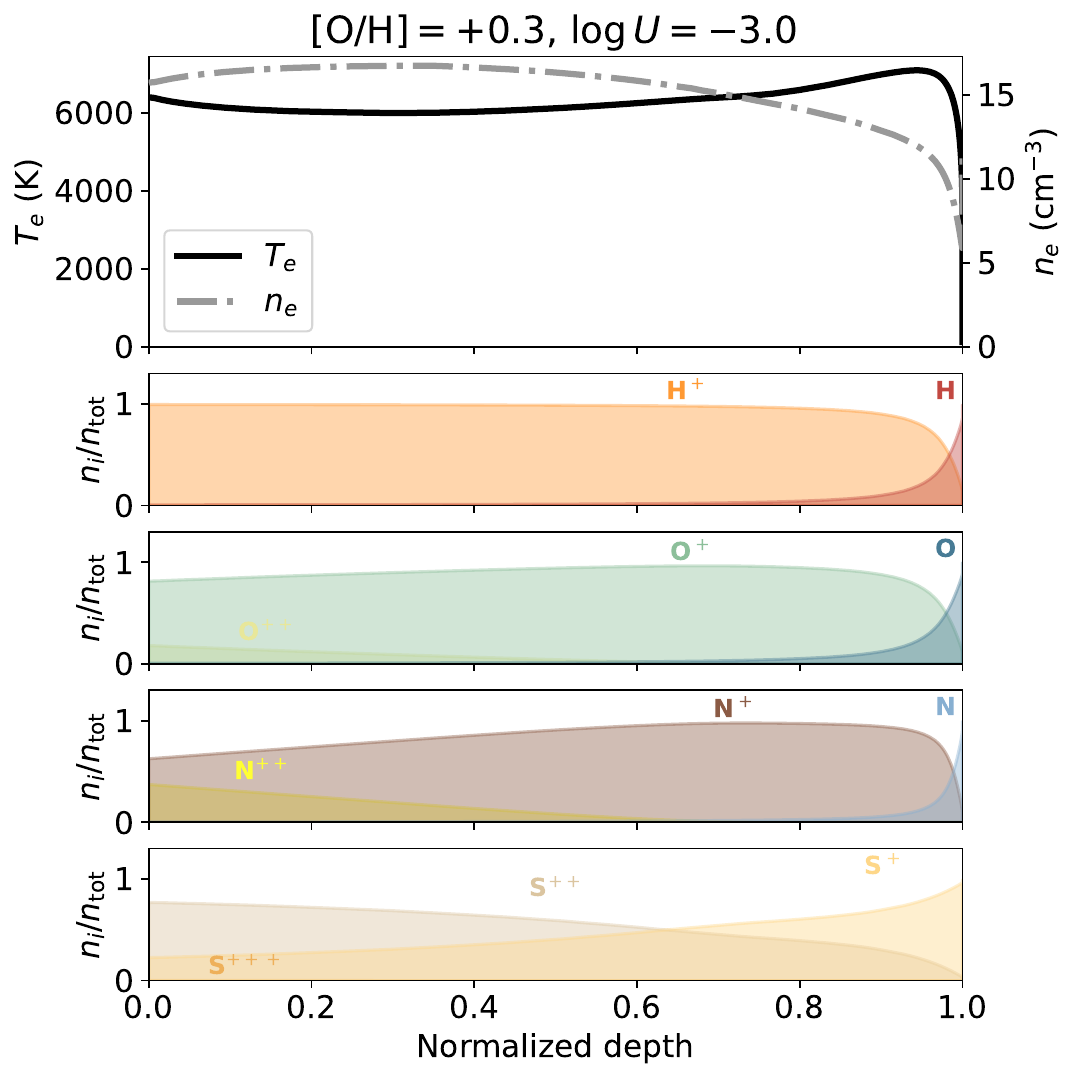}
\caption{Temperature, density, and ionization structures in four representative photoionization models. The upper panels show $T_e$ and $n_e$ as functions of normalized depth; the lower panels show the ion fractions of H, O, N, and S. The correspondence between the thermal and ionization gradients illustrates why different ionic phases sample different temperature distributions.}
\label{fig:temperature_profiles}
\end{figure*}

Figure~\ref{fig:temperature_profiles} shows strong thermal variations near the ionization front, where the filtered radiation field changes rapidly as hydrogen becomes neutral. In the constant-pressure calculations, the electron density also responds to the temperature gradient. Most importantly, this thermal structure coincides with changes in ionic composition: \Opp\ dominates farther into the highly ionized gas, whereas \Op\ and \Np\ extend toward the ionization front. Temperature diagnostics weighted by \Opp\ and \Np\ therefore sample different thermal structures and can yield systematically different temperatures even in the absence of additional small-scale temperature fluctuations.

The classical manifestation of this stratification is the $T_2$--$T_3$ relation. Here $T_2$ denotes a low-ionization temperature, commonly $T_e(\Nii)$ or $T_e(\Oii)$, whereas $T_3$ denotes $T_e(\Oiii)$. Systematic differences between these temperatures were recognized in early photoionization calculations \citep{stasinska82,campbell86,garnett92}. Because \Np\ and \Op\ have broadly similar ionization structures, either can serve as a tracer of the low-ionization temperature, although recent observations show that $T_e(\Oii)$ and $T_e(\Nii)$ may not be identical in detail \citep[e.g.,][]{mendez-delgado23b,rickards-vaught24}.

A commonly adopted relation of \citet{campbell86} and \citet{garnett92} is
\begin{equation}
T_2=0.7 \times T_3+3000~{\rm K}.
\label{eq:t2-t3-classical}
\end{equation}
For $T_3>10^4$ K, this relation predicts $T_3>T_2$, with the difference increasing toward higher temperature. Below $10^4$ K, the two temperatures become comparable, and $T_2$ can exceed $T_3$. This behavior arises naturally from the changing balance of heating and cooling across the nebula and does not require an additional stochastic temperature component.

\begin{figure*}[ht!]
\centering
\includegraphics[width=0.9\textwidth]{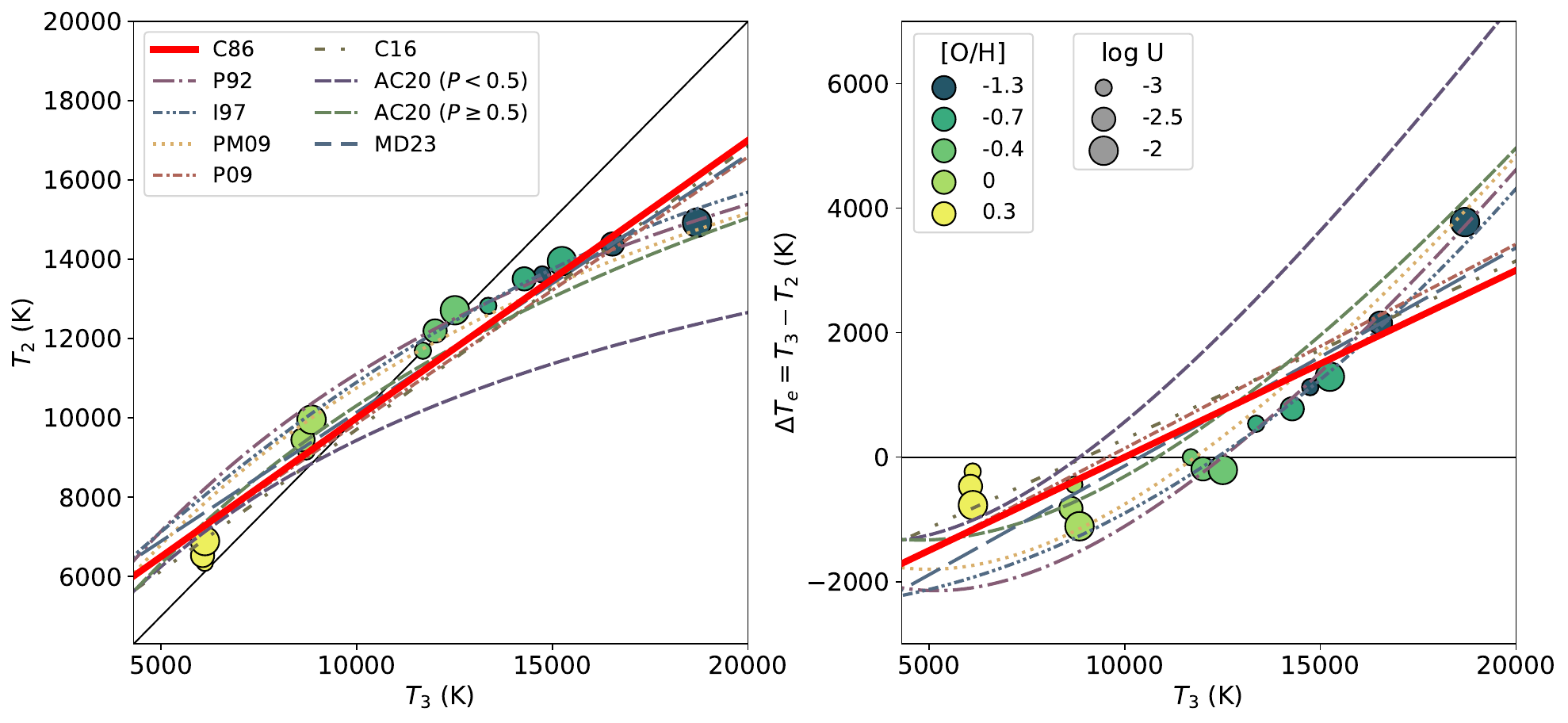}
\caption{Summary of literature $T_2$--$T_3$ relations. \textit{Left:} $T_2$ as a function of $T_3$. Circles represent the photoionization grid, with color indicating metallicity and point size indicating $\log U$. Curves represent P92 \citep{pagel92}, I97 \citep{izotov97}, PM09 \citep{perez-montero09}, P09 \citep{pilyugin09}, C16 \citep{croxall16}, AC20 \citep{arellano-cordova20}, and MD23 \citep{mendez-delgado23b} relations; the fiducial C86 relation \citep{campbell86} is shown in red. \textit{Right:} The same relations expressed as $\Delta T_e=T_3-T_2$ versus $T_3$.}
\label{fig:t2_t3}
\end{figure*}

Alternative relations have been derived from later photoionization models or calibrated against observations \citep[e.g.,][]{pagel92,oey00,ercolano07,perez-montero09,pilyugin09,bresolin09,esteban09,berg15,croxall16}. Some introduce a second excitation parameter ($P$) to account for variations in ionization state at fixed temperature \citep{pilyugin07,arellano-cordova20}. As shown in Figure~\ref{fig:t2_t3}, these prescriptions differ in curvature and normalization, while the photoionization models presented above exhibit additional scatter with metallicity and ionization parameter. Their common first-order behavior is nevertheless an increase in $T_3-T_2$ with $T_3$ over the temperature range occupied by metal-poor \Hii\ regions. Equation~\ref{eq:t2-t3-classical} therefore provides a useful baseline for the calculation below.

\subsection{The Conversion between the ADF and $t^2$}
\label{sec:adf-to-t2}

The fluctuation parameter is a dimensionless variance but is generally not observed directly. Instead, it is inferred by requiring a pair $(T_0,t^2)$ to simultaneously reproduce the temperatures and abundances implied by CEL and RL fluxes. This conversion is nonlinear because it depends on the derivatives of the emissivity functions shown in Figure~\ref{fig:emisivity}.

Expanding each emissivity to second order around $T_0$, \citet{peimbert13} expressed the apparent temperature inferred from two lines as
\begin{equation}
T_e(\lambda_1/\lambda_2)=T_0\left[
1+T_0\left(
\frac{\zeta_{\lambda_1}-\zeta_{\lambda_2}}
{\eta_{\lambda_1}-\eta_{\lambda_2}}
\right)\frac{t^2}{2}\right],
\label{eq:general-peimbert}
\end{equation}
where $\eta_\lambda=\varepsilon'_\lambda/\varepsilon_\lambda$ and
$\zeta_\lambda=\varepsilon''_\lambda/\varepsilon_\lambda$. Substituting the CEL and RL emissivity forms from Equations~\ref{eq:cel_emissivity} and \ref{eq:rl_emissivity} gives
\begin{eqnarray}
T_e(4363/5007)
&=&T_0\left[1+
\left(\frac{91300}{T_0}-2.68\right)\frac{t^2}{2}\right],
\label{eq:4363}\\[1ex]
T_e({\rm V1}/5007)
&=&T_0\Biggl[1+
\Biggl(\frac{29160}{T_0}-3.095\nonumber\\
&&+\frac{0.415}{29160/T_0+0.415}\Biggr)\frac{t^2}{2}\Biggr].
\label{eq:v1}
\end{eqnarray}
The dust-corrected fluxes of \Oiii\ $\lambda4363$, \Oiii\ $\lambda5007$, and the summed \Oiirec\ V1 multiplet provide two independent line ratios, allowing $T_0$ and $t^2$ to be solved simultaneously.

\begin{figure}[ht!]
\centering
\includegraphics[width=0.95\linewidth]{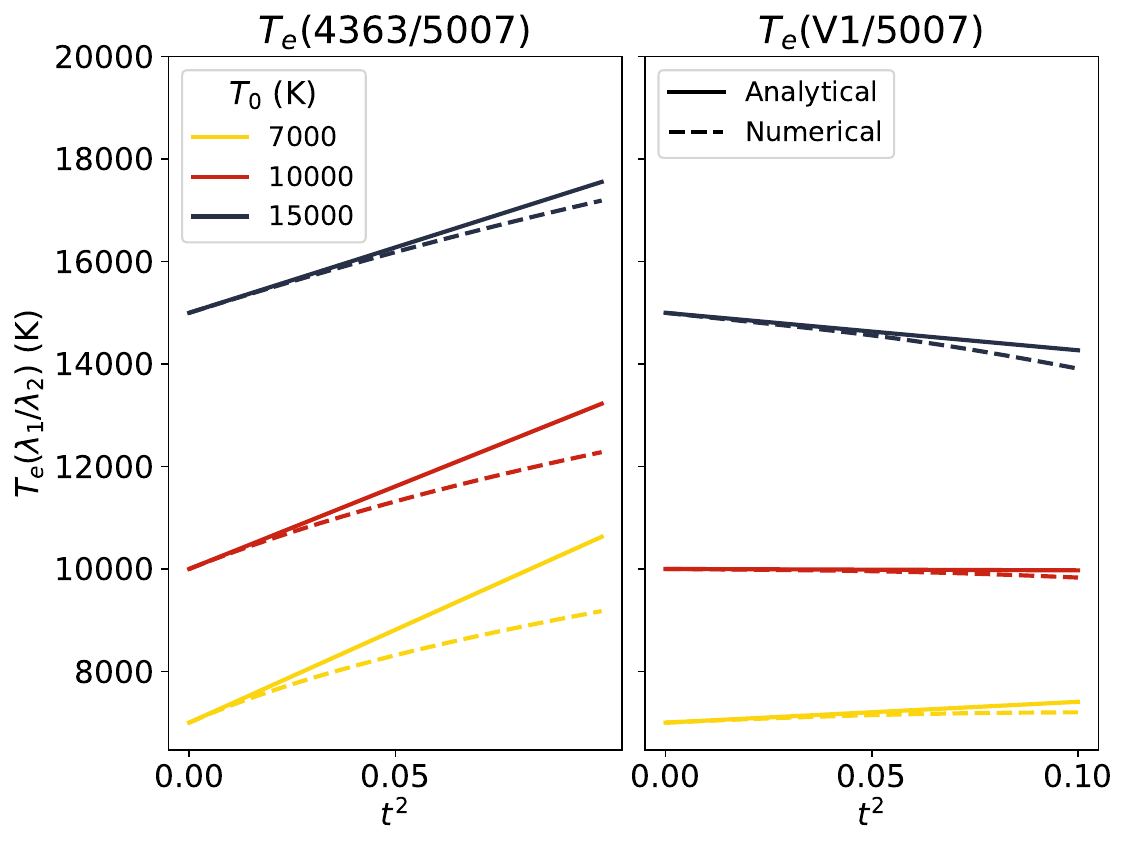}
\caption{Apparent $T_e(4363/5007)$ (left) and $T_e({\rm V1}/5007)$ (right) as functions of $t^2$ for different mean temperatures $T_0$. Solid lines show the second-order analytical expressions from Equations~\ref{eq:4363} and \ref{eq:v1}; dashed curves show direct numerical integrations over Gaussian temperature distributions detailed in Appendix \ref{apx:numerical-method}.}
\label{fig:Te_t2}
\end{figure}

Figure~\ref{fig:Te_t2} illustrates the behavior of the two apparent temperatures. At fixed $T_0$, increasing $t^2$ raises $T_e(4363/5007)$ substantially because the \Oiii\ auroral line is strongly weighted toward the hot tail of the temperature distribution. By comparison, $T_e({\rm V1}/5007)$ changes more slowly, so the separation between the two temperatures encodes the inferred fluctuation amplitude. The analytical expressions in Equations \ref{eq:general-peimbert}--\ref{eq:v1} are second-order Taylor expansions and are therefore most accurate near $t^2=0$. At larger variance, neglected higher-order terms become important. I therefore also calculate the apparent line fluxes by numerically integrating the emissivities over a Gaussian temperature distribution, following \citet{chen23,chen26} and described in detail in Appendix~\ref{apx:numerical-method}. At large $t^2$, the numerical calculation predicts a weaker response of the line ratios to increasing variance. Consequently, the analytical method infers a smaller $t^2$ from the same set of line ratios.

The same calculation determines the abundances that would be inferred from the line ratios. I use the apparent $T_e(4363/5007)$ with \Oiii\ $\lambda5007/{\rm H}\beta$ to derive the CEL abundance, $12+\log(\Opp/\Hp)_{\rm CEL}$. The \Oiirec\ V1/H$\beta$ ratio similarly yields the RL abundance, $12+\log(\Opp/\Hp)_{\rm RL}$. Because V1 and H$\beta$ have similar temperature dependences, the inferred RL abundance depends only weakly on the adopted temperature; I use $T_e({\rm V1}/5007)$ for consistency. Each $(T_0,t^2)$ pair therefore maps to a unique apparent ADF within the assumed temperature-distribution model. The RL abundance remains close to the input abundance, while the CEL abundance is biased low by the temperature inferred from the auroral line.

\begin{figure*}[ht!]
\centering
\includegraphics[height=3in]{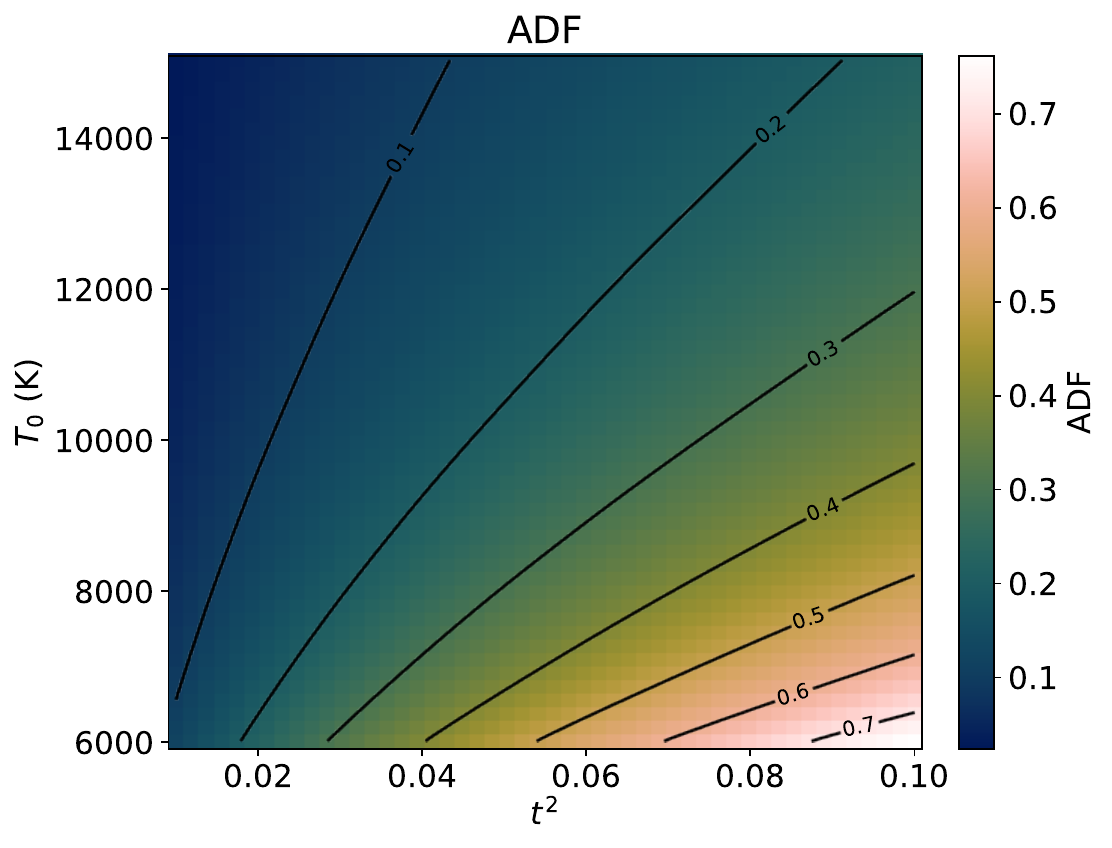}
\includegraphics[height=3in]{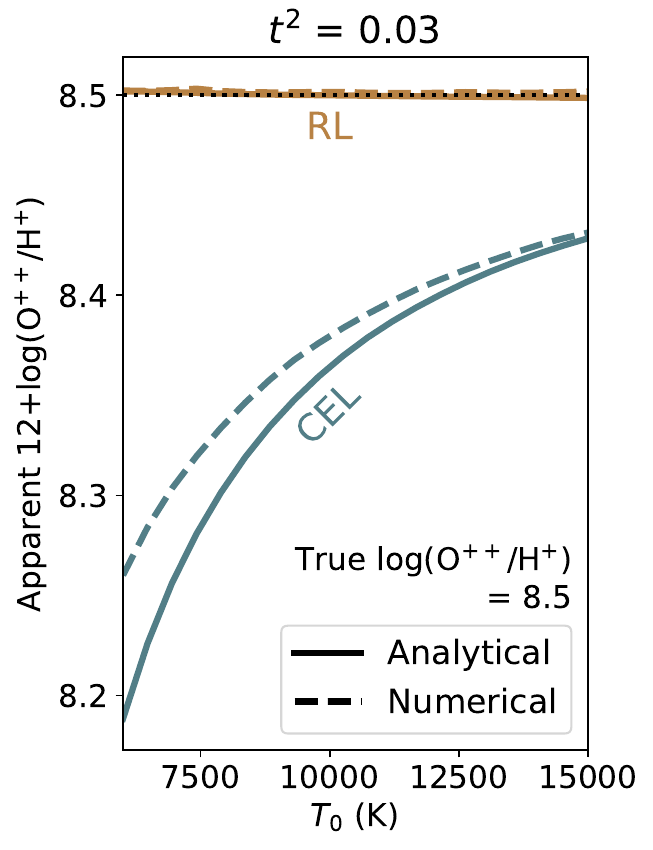}
\caption{\textit{Left:} Analytical ADF as a function of $T_0$ and $t^2$, calculated from Equations~\ref{eq:4363} and \ref{eq:v1}. \textit{Right:} Apparent CEL and RL \Opp/\Hp\ abundances as functions of $T_0$ for fixed $t^2=0.03$ and an input abundance of $12+\log(\Opp/\Hp)=8.5$. The separation between the inferred CEL and RL abundances defines the ADF, which varies with $T_0$ even at fixed $t^2$.}
\label{fig:adf_T0_t2}
\end{figure*}

Figure~\ref{fig:adf_T0_t2} shows that this mapping is strongly nonlinear. At $T_0=6000$ K, an ADF of approximately 0.2 dex is produced by only $t^2\simeq0.02$ in the analytical calculation, whereas the same ADF at $T_0=14000$ K requires approximately $t^2\simeq0.08$. This difference arises from the curvature of the optical CEL emissivity functions, which respond more strongly to a given fractional temperature perturbation at low $T_0$ than at high $T_0$. Thus, even if the ADF were identical in every nebula, the $t^2$ inferred from it would increase systematically with characteristic temperature.

\subsection{Constructing the Apparent $t^2$--$\Delta T_e$ Relation}
\label{sec:construct-correlation}

The preceding results suggest a direct null experiment. I combine two empirical properties without assuming that the interionic temperature difference directly traces temperature fluctuations within the \Opp\ zone:
\begin{enumerate}
\item the \Opp\ ADF is approximately constant across the current \Hii-region compilation; and
\item $T_e(\Oiii)-T_e(\Nii)$ follows the large-scale temperature structure represented by the $T_2$--$T_3$ relation.
\end{enumerate}

\begin{figure}[ht!]
\centering
\includegraphics[width=0.9\linewidth]{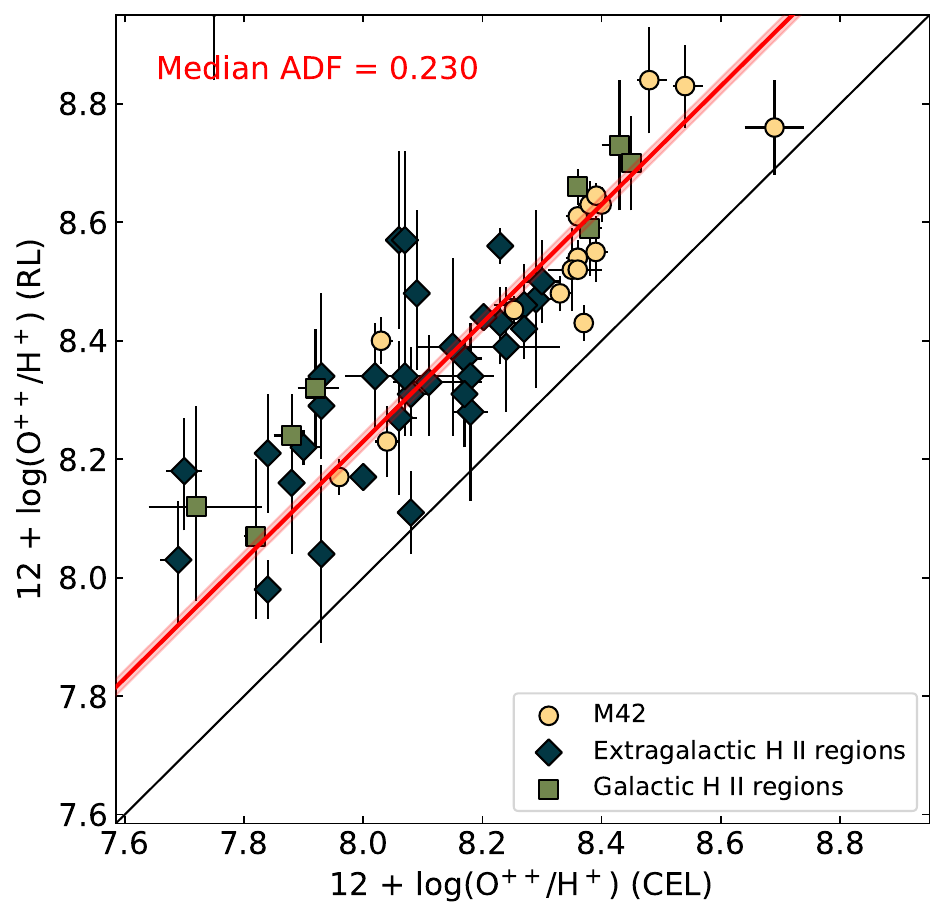}
\caption{Literature compilation of \Opp\ ADF measurements grouped by object category. The red line marks the sample median, ${\rm ADF}=0.230$ dex. Individual measurements and references are listed in Appendix~\ref{apx:target-summary}.}
\label{fig:adf_summary}
\end{figure}

Figure~\ref{fig:adf_summary} compiles literature measurements from \citet{mendez-delgado23nature,esteban09,esteban14,toribio-san-cipriano16,tsamis03,blagrave06,mendez-delgado21a,mendez-delgado21b,chen26,chen26orion}. The sample comprises 33 extragalactic \Hii\ regions, eight Galactic \Hii\ regions outside M42, and 18 measurements of distinct regions or kinematic components within M42. Individual entries are listed in Appendix~\ref{apx:target-summary}. The median ADF is $0.230\pm0.014$ dex, with no obvious dependence on object category or \Opp/\Hp\ over the range spanned by the sample.

Within the sample, Mrk~1259 is a notable outlier, with an ADF of 1.22 dex. It is a Wolf--Rayet galaxy with reported indications of possible active galactic nucleus (AGN) activity \citep{ohyama97}. The analytical calculation gives $t^2=0.143\pm0.018$, whereas the numerical Gaussian model requires $t^2>0.5$. Such a broad temperature distribution lies outside the regime in which a simple one-component Gaussian fluctuation model is physically informative. I therefore exclude Mrk~1259 from the subsequent analyses.

For the second structural component, I adopt Equation~\ref{eq:t2-t3-classical}. Over the temperature range spanned by the compiled sample, this relation provides a convenient approximation to the overall observed and modeled trend and is broadly consistent with the relation between \Nii- and \Oii-based low-ionization temperatures. The choice is not intended to imply zero intrinsic scatter. Rather, it defines a simple one-parameter baseline whose effect can be followed analytically.

For each value of $T_e(\Oiii)$, I first solve for the $(T_0,t^2)$ pair that produces ${\rm ADF}=0.23$ using the analytical formalism adopted in most of the literature compilation. Then the same $T_e(\Oiii)$ is mapped to $T_e(\Nii)$ using Equation~\ref{eq:t2-t3-classical}, yielding $\Delta T_e=T_e(\Oiii)-T_e(\Nii)$. Repeating this procedure constructs the predicted $t^2$--$\Delta T_e$ relation. Appendix~\ref{apx:numerical-method} repeats the calculation using direct numerical integration, which changes the quantitative values but does not alter the qualitative conclusions.

\begin{figure*}[ht!]
\centering
\includegraphics[height=3in]{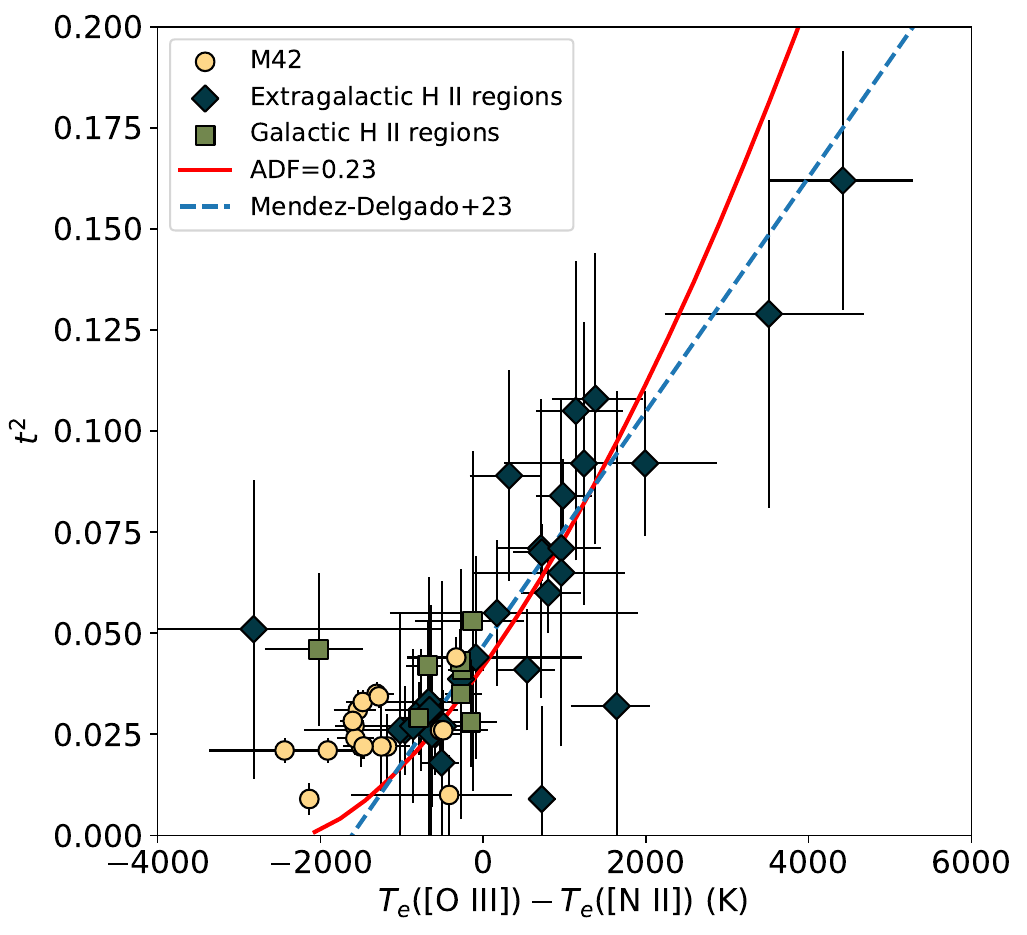}
\includegraphics[height=3in]{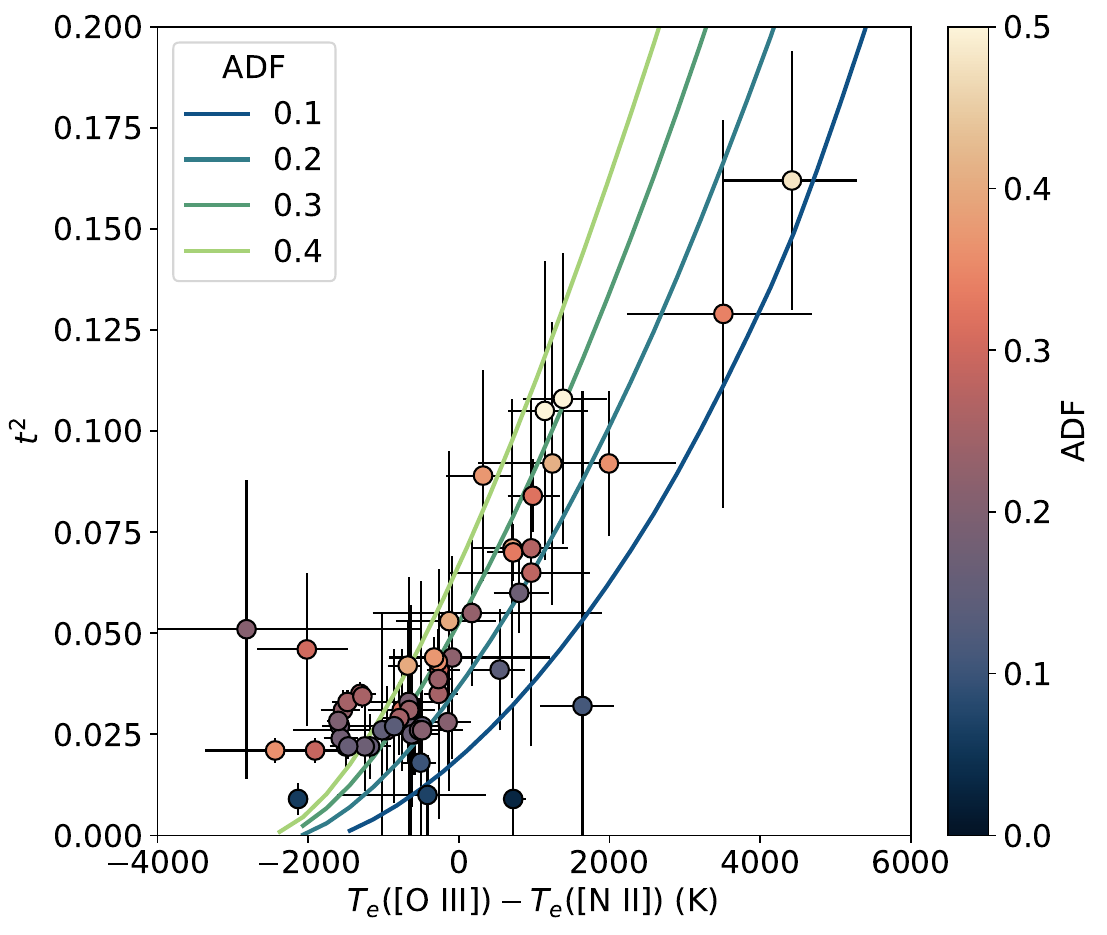}
\caption{
Predicted $t^2$--$\Delta T_e$ relation obtained by combining an approximately constant ADF with the classical $T_2$--$T_3$ relation, where $\Delta T_e=T_e(\Oiii)-T_e(\Nii)$. \textit{Left:} Literature measurements for M42, extragalactic \Hii\ regions, and other Galactic \Hii\ regions. The red curve assumes ${\rm ADF}=0.23$, while the dashed blue line shows the empirical fit of \citet{mendez-delgado23nature}. \textit{Right:} The same measurements color-coded by their individual ADFs, together with predicted curves for several fixed ADF values.}
\label{fig:t2_T3T2_summary}
\end{figure*}

As shown in Figure~\ref{fig:t2_T3T2_summary}, the constant-ADF prediction closely follows the observed positive correlation. This agreement results from the combination of two monotonic mappings. First, the $T_2$--$T_3$ relation causes $\Delta T_e$ to increase with $T_e(\Oiii)$. Second, at higher temperature, a larger $t^2$ is required to produce the same ADF. Thus, the horizontal coordinate is closely linked to the absolute temperature, while the vertical coordinate inherits a temperature dependence through the ADF-to-$t^2$ conversion. The two quantities can therefore correlate even without requiring $\Delta T_e$ to trace temperature fluctuations within the \Opp\ zone.

The right panel of Figure~\ref{fig:t2_T3T2_summary} provides a further consistency check. Objects with larger ADFs tend to lie above or toward the upper-left of the fiducial ADF$=0.23$ curve, whereas objects with smaller ADFs preferentially lie below or toward the lower-right. Curves calculated for several fixed ADFs reproduce this ordering. The observed scatter is therefore structured with respect to the same quantity from which $t^2$ is inferred.

These results show that a positive $t^2$--$\Delta T_e$ relation is expected when the ADF is approximately constant and the interionic temperatures follow a structural $T_2$--$T_3$ sequence. The correlation therefore does not by itself establish that temperature differences \textit{between} ionization zones trace temperature fluctuations \textit{within} the \Opp\ zone. Any mechanism that produces a similar ADF distribution will generate a comparable apparent trend when subjected to the same ADF-to-$t^2$ transformation.

\section{Discussion}
\label{sec:discussion}

\subsection{Structural Variations and Fluctuations within Ionic Zones}
\label{sec:structural-versus-local}

In \S\ref{sec:structural-origin}, I demonstrated that the observed $t^2$--$\Delta T_e$ correlation does not require the interionic temperature difference to trace the fluctuations responsible for the ADF. Therefore, the correlation  does not necessarily establish causation \citep{pearl16}. This does not imply that temperature fluctuations are absent. Their amplitude could, for example, vary among \Hii\ regions in such a way that the resulting ADF remains near 0.23 dex. More generally, multiple mechanisms could contribute in different proportions while producing a relatively narrow ADF distribution in an unresolved sample.

M42 provides the most direct evidence that this possibility must be evaluated object by object. The regions included in the compilation span compact apertures and kinematically resolved Herbig--Haro components, yet their \Oiirec\ excesses can be accounted for by continuum fluorescence at levels comparable to their measured ADFs \citep{chen26orion}. This result argues against interpreting the M42 measurements as direct evidence for large temperature fluctuations. However, I caution that the result should not be generalized to the full compilation without additional observations. The M42 points occupy a restricted region of Figure~\ref{fig:t2_T3T2_summary}, whereas metal-poor extragalactic \Hii\ regions extend to higher $T_e$, larger inferred $t^2$, and larger $\Delta T_e$. Continuum pumping, temperature structure, and other processes may therefore contribute in different proportions under physical conditions not sampled by M42.

\begin{figure*}[ht!]
\centering
\includegraphics[width=0.7\textwidth]{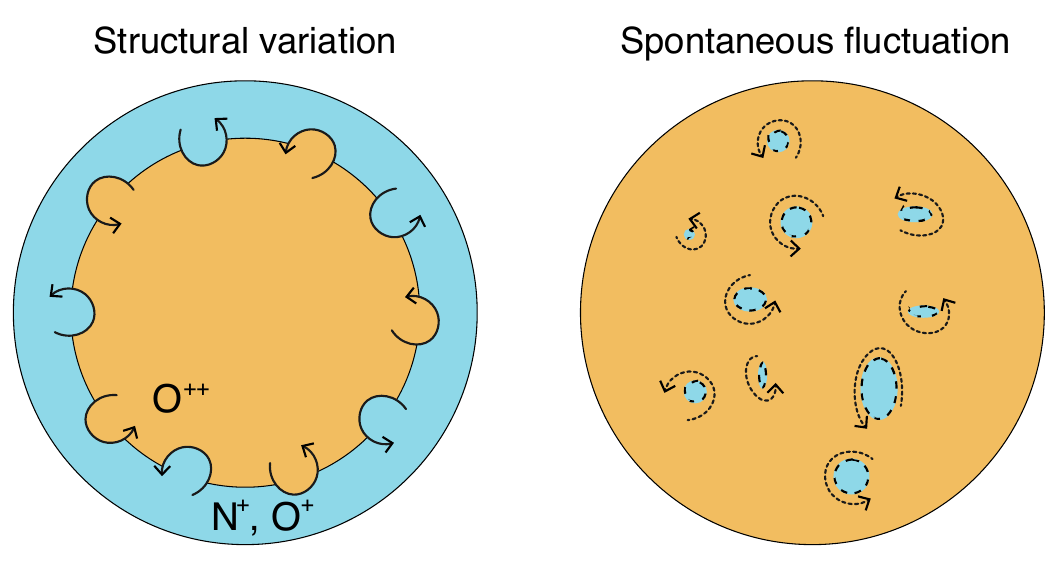}
\caption{Schematic comparison of two forms of temperature structure. \textit{Left:} Large-scale structural variation associated with ionization stratification. \textit{Right:} Small-scale temperature fluctuations within individual ionic phases. Yellow represents the high-ionization \Opp\ zone; cyan represents the lower-ionization \Np, \Op\ zone.}
\label{fig:structure_vs_spontaneous}
\end{figure*}

The distinction relevant to this work is illustrated in Figure~\ref{fig:structure_vs_spontaneous}. In the structural picture, the mean temperature changes systematically between the high-ionization \Opp\ zone and the lower-ionization \Np/\Op\ zone. This variation follows the large-scale ionization structure and is reproduced by equilibrium photoionization calculations. It therefore determines the interionic temperature difference without, by itself, implying a large temperature variance within either ionic zone.

By contrast, the fluctuation picture requires substantial temperature variations to be distributed throughout the \Opp-emitting gas. If the ADF is caused primarily by such variations and its inferred $t^2$ is physically connected to $\Delta T_e$, then the processes that establish the large-scale ionization gradient must also regulate, or at least correlate with, the small-scale thermal structure within the \Opp\ zone. Turbulence, shocks, magnetic processes, cosmic rays, or unresolved density structure \citep{giammanco05,mendez-delgado23b} could in principle introduce additional heating and cooling variations. The key physical requirement is not merely that such processes exist, but that the resulting ion-weighted temperature variance systematically track the temperature separation between distinct ionization zones.

The spatial scales probed by existing observations further emphasize this distinction. In M42, variations associated with the ADF have been examined down to projected scales of approximately 0.002 pc \citep{chen26orion,mesa-delgado08}, while measurements in other Galactic \Hii\ regions probe scales of approximately 0.007--0.035 pc \citep{mesa-delgado10}. Line-of-sight mixing remains unavoidable, but a genuine physical link between $t^2(\Opp)$ and $\Delta T_e$ would require the small-scale temperature distribution sampled by \Opp\ emission to remain coupled to the larger-scale thermal structure separating the low- and high-ionization zones. Standard one-dimensional photoionization models naturally generate the latter structure but do not generate the former at the amplitude inferred from large ADFs.

\subsection{Comparison with Previous Photoionization Models}
\label{sec:previous-models}

The distinction between ionization structure and the origin of the ADF is precisely consistent with detailed modeling of individual nebulae. H1013, a giant \Hii\ region in M101 included in the present compilation, was modeled by \citet{stasinska13} for optical and mid-infrared CELs spanning a broad range of excitation energies and temperature sensitivities. Their photoionization model matched the principal CEL constraints without invoking additional temperature fluctuations but did not account for the enhanced \Oiirec\ RL emission. The model could therefore explain the thermal structure diagnosed by the CELs while leaving the ADF unresolved. This is precisely the distinction emphasized here: reproducing interionic and CEL temperature structure does not by itself identify the process responsible for excess RL emission.

Earlier model grids reached a related conclusion from a different direction. \citet{garnett92} found that temperature variations within the \Opp\ zone produce a substantial bias in the optical CEL temperature primarily in cool, metal-rich nebulae, approximately $T_e\lesssim5000$ K. At high metallicity, efficient infrared fine-structure cooling can generate steep internal temperature gradients, causing the optical lines to preferentially weight the warmer parts of the \Opp\ zone. This behavior differs from the observed $t^2$--$\Delta T_e$ sequence, in which the largest inferred $t^2$ values occur predominantly in hotter, metal-poor objects. The modeled trend therefore runs opposite to the observed $t^2$--$\Delta T_e$ relation. 

\citet{ercolano07} investigated whether inhomogeneous gas distributions and multiple ionizing sources could increase the temperature variance. Their models reached $t^2$ values near 0.1 only at high metallicity ($\sim$ twice solar). Geometry and source distribution can therefore enhance temperature structure, but the largest modeled effect again occurs in a different regime from the high-temperature end of the observed relation. Taken together, these calculations indicate that the process responsible for the ADF should not be assumed to be the same process that establishes the ordinary interionic temperature gradient.

\subsection{Observational Tests with Greater Discriminating Power}
\label{sec:future-tests}

Following the preceding analysis, here I suggest four practical observational tests for determining whether temperature structure within individual ionic zones contributes to the ADF and for distinguishing this explanation from alternative mechanisms.

\begin{enumerate}
\item \textit{Begin with the measured quantities.} Correlations should first be examined using the ADF or the underlying dust-corrected flux ratios, rather than only after conversion to $t^2$. The ADF-to-$t^2$ mapping is nonlinear and depends strongly on absolute temperature. Comparing the untransformed and transformed planes can therefore reveal trends introduced by the conversion itself.

\item \textit{Choose a diagnostic pair that samples overlapping gas.} To probe temperature variations within the \Opp\ zone, compare diagnostics that sample broadly overlapping volumes and weight hot and cool gas differently. The classical  $T_e(\Oiii)-T_e({\rm BJ})$ relation provide one such pair, although hydrogen recombination emission also extends beyond the \Opp\ zone. For this pair, \citet{garcia-rojas07} found a shallower relation between the ADF and the temperature difference in
\Hii\ regions than in planetary nebulae. 

\item \textit{Test the model against additional diagnostics.} Fit a temperature-distribution model to one diagnostic pair, then test whether it also predicts independent line ratios. \ion{He}{1} ratios can provide such a test if the model accounts for the different emitting volumes of \Hep\ and \Opp, as well as density sensitivity, metastable-level populations, optical-depth effects, and underlying stellar absorption.

\item \textit{Test the relation spatially within individual nebulae.} Global $T_2$--$T_3$ relations are calibrated by comparing integrated \Hii\ regions. Within a resolved nebula, local variations in density, ionization parameter, geometry, and the filtered radiation field can help distinguish structural temperature differences from a fluctuation field distributed throughout the \Opp\ gas. The breakdown of the global $t^2$--$\Delta T_e$ relation among resolved M42 components \citep{chen26orion} is therefore informative, but equivalent measurements across other Galactic and nearby extragalactic \Hii\ regions are needed to determine whether this behavior is general.
\end{enumerate}

Together, these tests provide greater discriminating power among temperature fluctuations, chemically distinct gas, continuum pumping, and other mechanisms that alter the CEL--RL abundance comparison.

\section{Summary}
\label{sec:summary}

I have reassessed the interpretation of the observed relation between the temperature-fluctuation parameter $t^2(\Opp)$ and the interionic temperature difference $\Delta T_e=T_e(\Oiii)-T_e(\Nii)$ in star-forming \Hii\ regions. The analysis combines photoionization models, analytical and numerical temperature-distribution calculations, and a compilation of 59 integrated, spatially resolved, and kinematically resolved measurements.

The photoionization models demonstrate that interionic temperature differences arise naturally from ionization stratification. The high- and low-ionization zones sample different portions of the nebular thermal structure and consequently follow a systematic $T_2$--$T_3$ relation without requiring an additional stochastic temperature field within the \Opp\ zone. Independently, the conversion between the ADF and $t^2$ is nonlinear and strongly dependent on absolute temperature. The compiled sample has a median ADF of $0.230\pm0.014$ dex, with no clear dependence on object category or \Opp/\Hp\ over the range represented here. Combining this approximately constant ADF with the classical structural $T_2$--$T_3$ relation produces an apparent $t^2$--$[T_e(\Oiii)-T_e(\Nii)]$ sequence that closely follows the observed correlation.

The observed correlation therefore does not require $T_e(\Oiii)-T_e(\Nii)$ to trace the temperature fluctuations responsible for the ADF. A comparable trend is expected for any mechanism that produces an approximately constant ADF when subjected to the same temperature-dependent ADF-to-$t^2$ transformation. This conclusion does not demonstrate that temperature fluctuations are absent, nor does it identify a single universal alternative explanation. Continuum pumping accounts for the available M42 measurements, but its importance across the full range of metallicity and excitation remains to be determined. Temperature fluctuations, chemically distinct gas, fluorescence, and other mechanisms may contribute in different proportions among \Hii\ regions.

Thus, the principal implication concerns the diagnostic power of the $t^2$--$\Delta T_e$ correlation: by itself, it is not a unique test of temperature fluctuations within the \Opp\ phase. More discriminating tests should compare the primary line ratios before transformation, match the ionic and spatial weighting of the temperature diagnostics, require a single physical model to reproduce several independent temperature diagnostics, and repeat these comparisons with spatially resolved observations. The Sloan Digital Sky Survey Local Volume Mapper (SDSS/LVM) \citep{drory24} and the ongoing Comprehensive IFU Research on Accurate Abundance Studies (CIRAAS) project \citep{chen23,chen26,chen26orion} provide particularly suitable data sets for carrying out these tests. Subsequent work in this series will also examine other evidence commonly interpreted in support of temperature fluctuations.

\begin{acknowledgments}

Y.C. would like to thank Tucker Jones and Renbin Yan for the preliminary discussions about this analysis. Y.C. gratefully acknowledge support for this work from the National Aeronautics and Space Administration (NASA) under grant 80NSSC23K1132, the Direct Grant for Research (C0010-4053720, C001-4053782) from the Faculty of Science, The Chinese University of Hong Kong, and the National Natural Science Foundation of China (NSFC) under the Young Scientists Fund (Category C) 12603045. This work is supported by the Hong Kong Jockey Club Charities Trust through the project, JC STEM Lab of Astronomical Instrumentation.

\end{acknowledgments}

\begin{contribution}

Y.C. conceived the study, compiled the data, performed the modeling and analysis, prepared the figures, and wrote the manuscript.

\end{contribution}

\facilities{N.A.}

\software{ 
          Cloudy \citep{ferland13}, 
          }

\appendix

\section{Literature Sample}
\label{apx:target-summary}

Table~\ref{tab:target-summary} lists the 59 measurements used in the analysis. This sample contains all \Hii\ regions compiled in \citet{mendez-delgado23nature}. This sample contains 33 extragalactic \Hii\ regions (E--HII), eight Galactic \Hii\ regions outside M42 (G--HII), and 18 spatially or kinematically resolved measurements within M42. Multiple entries for the same named nebula are retained when they correspond to independent apertures, kinematic components, or literature measurements.

The temperature columns give $T_e(\Oiii)$ from the auroral-to-nebular \Oiii\ ratio and $T_e(\Nii)$ from the corresponding \Nii\ diagnostic. The ADF is defined by Equation~\ref{eq:adf}.

The columns $t^2_{\rm ana}$ and $t^2_{\rm num}$ report solutions obtained using the analytical and numerical methods described in Section~\ref{sec:adf-to-t2} and Appendix~\ref{apx:numerical-method}, respectively. Mrk~1259 requires $t^2>0.5$ in the numerical Gaussian model and is therefore reported as a lower limit. Reference numbers are expanded in the table notes.

\begingroup
\makeatletter
\patchcmd{\longdeluxetable}
  {\multicolumn{\totalcolumns}{c}{\vrule height 24pt width0pt\small\it
   \fnum@table\
   continued  \if@two@col\else on next page\fi}\\}
  {}{}{\PackageWarning{manuscript}{Could not suppress table continuation footer}}
\makeatother
\input{appendix_target_table}
\endgroup

\section{Numerical Temperature-Distribution Calculation}
\label{apx:numerical-method}

The analytical relations in Equations~\ref{eq:4363} and \ref{eq:v1} approximate each emissivity by a second-order Taylor expansion about $T_0$. To assess the effect of neglected higher-order terms, I repeat the calculation by integrating the \textsc{PyNeb} emissivities directly over a Gaussian temperature distribution. For a specified $(T_0,t^2)$, the standard deviation is
\begin{equation}
\sigma_T=T_0\sqrt{t^2},
\end{equation}
and the emissivity averaged over the \Opp\ volume is evaluated as
\begin{equation}
\left\langle\varepsilon_\lambda\right\rangle =
\frac{\int_{T_{\rm min}}^{T_{\rm max}}
\varepsilon_\lambda(T,n_e)
\exp[-(T-T_0)^2/(2\sigma_T^2)]\,dT}
{\int_{T_{\rm min}}^{T_{\rm max}}
\exp[-(T-T_0)^2/(2\sigma_T^2)]\,dT}.
\label{eq:numerical-emissivity}
\end{equation}
The integral is sampled at 1000 temperatures from $T_0-5\sigma_T$ to $T_0+5\sigma_T$, subject to a lower temperature floor of 500 K, and the retained Gaussian weights are renormalized. 

For each observed set of \Oiii\ $\lambda4363$, \Oiii\ $\lambda5007$, and \Oiirec\ V1 relative fluxes, I solve for the $(T_0,t^2)$ pair whose two independent line ratios reproduce the measurements. The model is restricted to $500\leq T_0\leq30000$ K and $0\leq t^2\leq0.5$. Solutions that reach the upper $t^2$ boundary without reproducing the flux ratios are reported as lower limits rather than finite measurements.

\begin{figure}[ht!]
\centering
\includegraphics[width=0.95\linewidth]{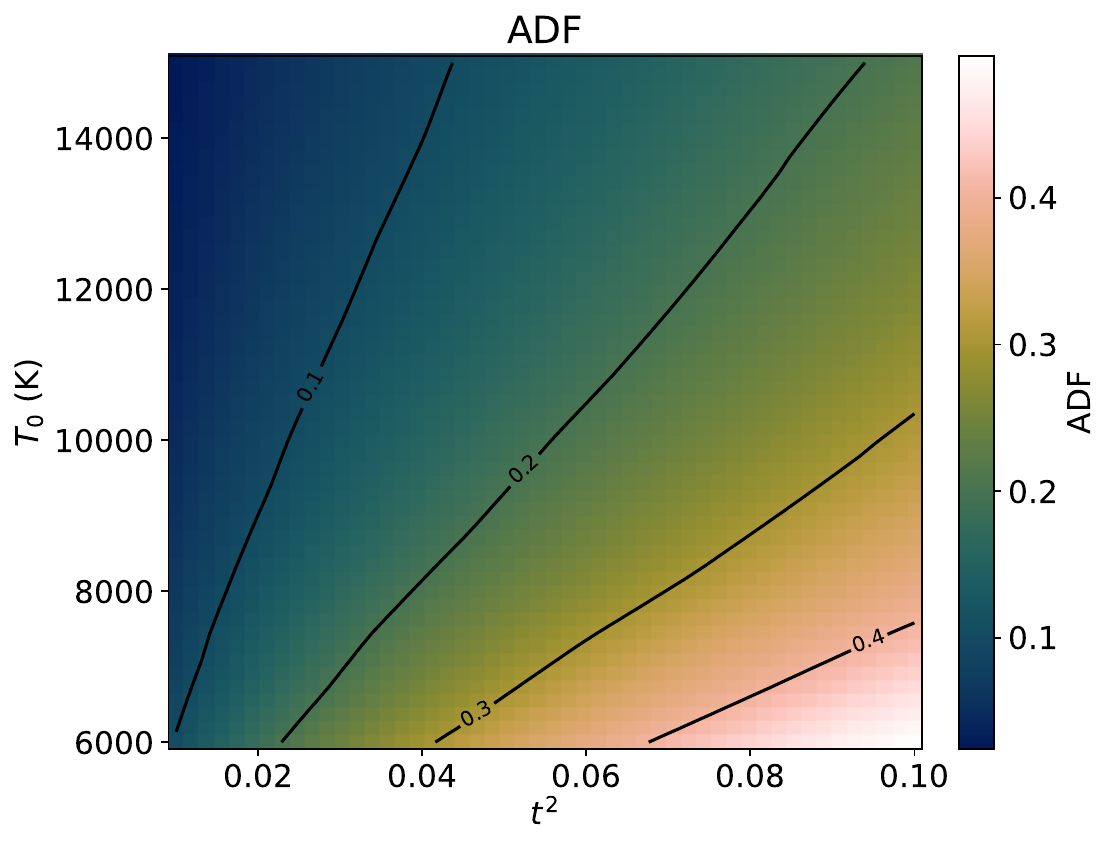}
\caption{Numerically calculated ADF as a function of $T_0$ and $t^2$, for comparison with the analytical surface in the left panel of Figure~\ref{fig:adf_T0_t2}. The response flattens at large $t^2$ as the width of the adopted Gaussian distribution becomes comparable to its mean temperature and the lower-temperature truncation becomes increasingly important. Within the parameter range explored here, the predicted ADF approaches $\sim$0.5 dex.
\label{fig:adf_T0_t2_numerical}}
\end{figure}

Figure~\ref{fig:adf_T0_t2_numerical} shows the numerical ADF surface. At small $t^2$, it approaches the analytical solution, as expected. At larger $t^2$, the second-order approximation increasingly overpredicts the response of the line ratios to the fluctuation amplitude. The analytical method therefore generally assigns a smaller $t^2$ than the numerical integration for the same ADF. Within the adopted Gaussian model, the temperature response flattens as $\sigma_T$ becomes a substantial fraction of $T_0$. The finite lower-temperature boundary and renormalization of the truncated distribution also become increasingly important in this regime. Consequently, the ADF within the explored grid approaches $\sim$0.5 dex rather than increasing indefinitely.

\begin{figure*}[ht!]
\centering
\includegraphics[height=3in]{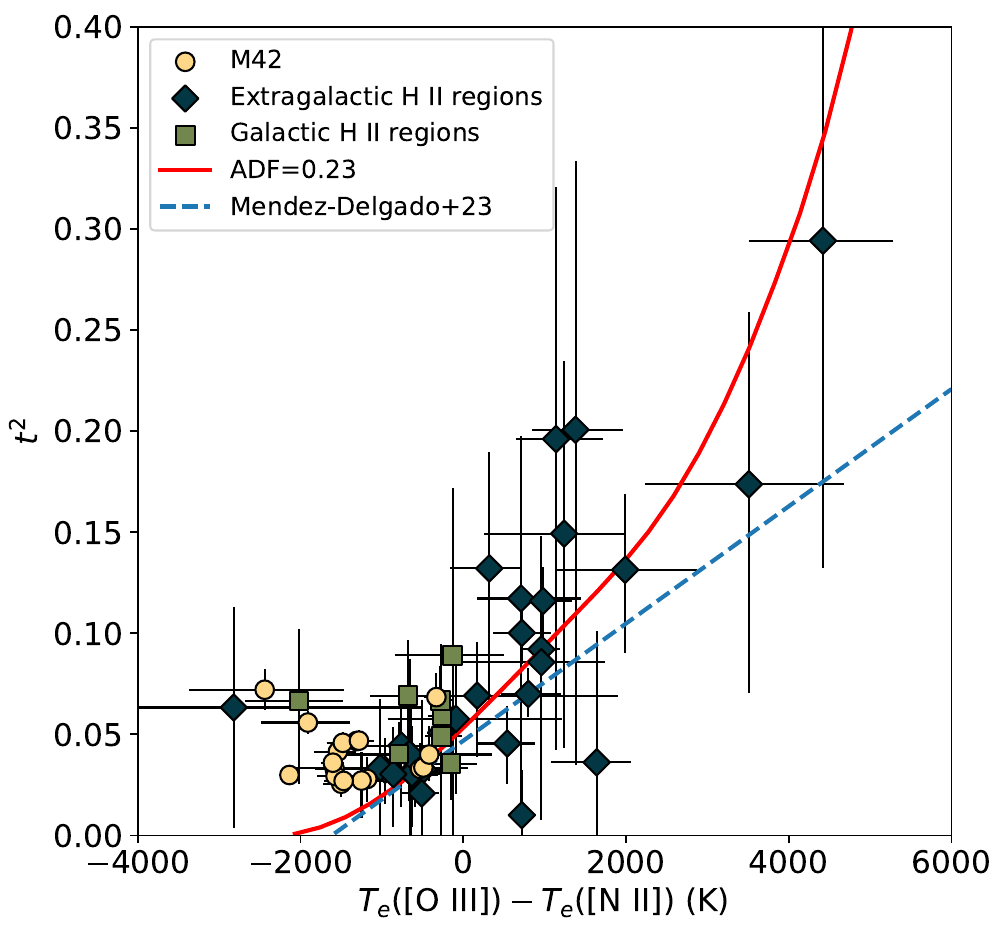}
\includegraphics[height=3in]{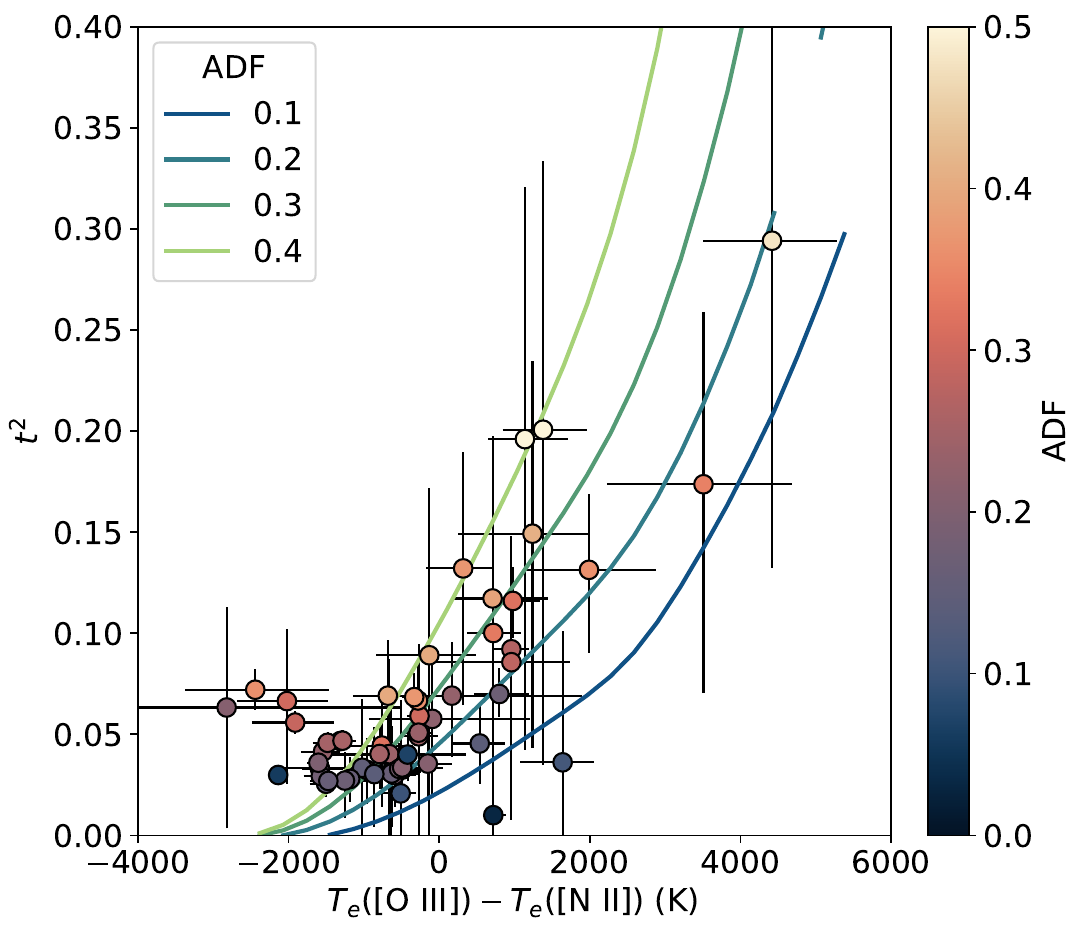}
\caption{Same as Figure~\ref{fig:t2_T3T2_summary}, but using the numerical temperature-distribution calculation for both the measured points and the fixed-ADF predictions. The numerical method yields larger $t^2$ values, particularly at large ADF, while preserving the close correspondence between the observations and the constant-ADF prediction.
\label{fig:t2_T3T2_summary_num}}
\end{figure*}

Figure~\ref{fig:t2_T3T2_summary_num} repeats the principal $t^2$--$\Delta T_e$ comparison using numerical $t^2$ values for both the median ADF and the fixed-ADF intervals. The numerical values are generally larger, and their uncertainties become more asymmetric at high $t^2$, but the observed points continue to follow the relation produced by an approximately constant ADF combined with the structural $T_2$--$T_3$ relation. Accordingly, the choice between the analytical and numerical solutions changes the inferred fluctuation amplitudes but not the interpretation of the correlation.

\input{PASPmanuscript_section_edit.bbl}
\end{document}

%% file: appendix_target_table.tex
\begin{longrotatetable}
\begin{deluxetable*}{llcccccc}
\tablecaption{Literature H II-region temperature and abundance-discrepancy compilation.\label{tab:target-summary}}
\tablewidth{0pt}
\tabletypesize{\scriptsize}
\tablehead{
\colhead{Object} & \colhead{Type} &
\colhead{$T_e$(\Oiii)} & \colhead{$T_e$(\Nii)} &
\colhead{ADF(\Opp)} & \colhead{$t^2_\mathrm{ana/lit}$(\Opp)} &
\colhead{$t^2_\mathrm{num}$(\Opp)} & \colhead{Ref.} \\
\colhead{} & \colhead{} & \colhead{(K)} & \colhead{(K)} &
\colhead{(dex)} & \colhead{} & \colhead{} & \colhead{}
}
\startdata
Haro 3 & E--HII & $9498_{-86}^{+95}$ & $9770_{-140}^{+150}$ & $0.238_{-0.031}^{+0.031}$ & $0.039\pm0.004$ & $0.051_{-0.008}^{+0.008}$ & 1 \\
IC132 & E--HII & $10280_{-540}^{+450}$ & $11300_{-980}^{+1050}$ & $0.149_{-0.142}^{+0.117}$ & $0.026\pm0.029$ & $0.033_{-0.041}^{+0.034}$ & 2 \\
IC2111 & E--HII & $9130_{-200}^{+180}$ & $9760_{-310}^{+360}$ & $0.161_{-0.098}^{+0.098}$ & $0.025\pm0.018$ & $0.031_{-0.027}^{+0.024}$ & 3 \\
LMC-30Dor & E--HII & $9870_{-40}^{+90}$ & $10360_{-240}^{+280}$ & $0.149_{-0.051}^{+0.051}$ & $0.027\pm0.009$ & $0.033_{-0.012}^{+0.012}$ & 4 \\
LMC-N11B & E--HII & $9140_{-90}^{+130}$ & $9780_{-150}^{+160}$ & $0.199_{-0.042}^{+0.032}$ & $0.031\pm0.006$ & $0.039_{-0.009}^{+0.009}$ & 3 \\
LMC-N44C & E--HII & $11280_{-140}^{+150}$ & $10560_{-330}^{+320}$ & $0.330_{-0.036}^{+0.036}$ & $0.070\pm0.007$ & $0.100_{-0.015}^{+0.015}$ & 3 \\
M31-K932 & E--HII & $8300_{-130}^{+140}$ & $9260_{-190}^{+170}$ & $0.199_{-0.076}^{+0.076}$ & $0.026\pm0.011$ & $0.032_{-0.017}^{+0.016}$ & 5 \\
Mrk1259 & E--HII & $9970_{-260}^{+460}$ & $7860_{-280}^{+250}$ & $1.220_{-0.143}^{+0.130}$ & $0.143\pm0.018$ & $>0.5$ & 6 \\
NGC1714 & E--HII & $9520_{-220}^{+170}$ & $10190_{-420}^{+560}$ & $0.190_{-0.076}^{+0.086}$ & $0.033\pm0.015$ & $0.040_{-0.023}^{+0.021}$ & 3 \\
NGC2363 & E--HII & $16480_{-390}^{+450}$ & $12970_{-1080}^{+1220}$ & $0.340_{-0.112}^{+0.104}$ & $0.129\pm0.048$ & $0.174_{-0.103}^{+0.085}$ & 5 \\
NGC2403-VS44 & E--HII & $8210_{-170}^{+240}$ & $8870_{-250}^{+330}$ & $0.241_{-0.155}^{+0.162}$ & $0.031\pm0.033$ & $0.040_{-0.059}^{+0.047}$ & 5 \\
NGC346 & E--HII & $12820_{-110}^{+120}$ & $10830_{-880}^{+860}$ & $0.360_{-0.071}^{+0.071}$ & $0.092\pm0.018$ & $0.131_{-0.041}^{+0.037}$ & 7 \\
NGC456-2 & E--HII & $12070_{-220}^{+170}$ & $11110_{-760}^{+1070}$ & $0.281_{-0.122}^{+0.130}$ & $0.065\pm0.043$ & $0.086_{-0.078}^{+0.062}$ & 8 \\
NGC456a-1 & E--HII & $12210_{-130}^{+70}$ & $11890_{-380}^{+470}$ & $0.369_{-0.100}^{+0.100}$ & $0.089\pm0.026$ & $0.132_{-0.068}^{+0.057}$ & 6 \\
NGC456a-2 & E--HII & $11940_{-100}^{+70}$ & $10700_{-760}^{+980}$ & $0.410_{-0.140}^{+0.140}$ & $0.092\pm0.035$ & $0.149_{-0.106}^{+0.085}$ & 6 \\
NGC5253-HII-1 & E--HII & $11950_{-160}^{+240}$ & $10810_{-520}^{+460}$ & $0.500_{-0.151}^{+0.151}$ & $0.105\pm0.037$ & $0.196_{-0.154}^{+0.125}$ & 9 \\
NGC5253-HII-2 & E--HII & $11900_{-190}^{+240}$ & $10520_{-530}^{+500}$ & $0.511_{-0.151}^{+0.151}$ & $0.108\pm0.036$ & $0.201_{-0.166}^{+0.133}$ & 9 \\
NGC5253-UV-1 & E--HII & $10840_{-130}^{+190}$ & $10130_{-710}^{+520}$ & $0.389_{-0.132}^{+0.141}$ & $0.071\pm0.037$ & $0.117_{-0.101}^{+0.080}$ & 9 \\
NGC5253C1 & E--HII & $12140_{-70}^{+80}$ & $11420_{-150}^{+100}$ & $0.029_{-0.071}^{+0.071}$ & $0.009\pm0.023$ & $0.010_{-0.025}^{+0.022}$ & 6 \\
NGC5253C2 & E--HII & $10150_{-50}^{+80}$ & $11010_{-350}^{+440}$ & $0.140_{-0.091}^{+0.091}$ & $0.027\pm0.019$ & $0.030_{-0.026}^{+0.023}$ & 6 \\
NGC5253P2 & E--HII & $12280_{-110}^{+120}$ & $11320_{-200}^{+220}$ & $0.270_{-0.100}^{+0.100}$ & $0.071\pm0.030$ & $0.092_{-0.058}^{+0.049}$ & 6 \\
NGC5408 & E--HII & $15910_{-460}^{+330}$ & $11490_{-800}^{+790}$ & $0.480_{-0.104}^{+0.095}$ & $0.162\pm0.032$ & $0.294_{-0.162}^{+0.136}$ & 10 \\
NGC5455 & E--HII & $9610_{-150}^{+120}$ & $10120_{-170}^{+200}$ & $0.100_{-0.153}^{+0.133}$ & $0.018\pm0.045$ & $0.021_{-0.062}^{+0.046}$ & 11 \\
NGC5461 & E--HII & $8460_{-220}^{+200}$ & $9100_{-350}^{+300}$ & $0.179_{-0.158}^{+0.158}$ & $0.027\pm0.030$ & $0.031_{-0.048}^{+0.039}$ & 5 \\
NGC5471 & E--HII & $13820_{-150}^{+190}$ & $12180_{-370}^{+540}$ & $0.111_{-0.151}^{+0.150}$ & $0.032\pm0.078$ & $0.036_{-0.089}^{+0.065}$ & 11 \\
NGC588 & E--HII & $10810_{-380}^{+500}$ & $10900_{-1200}^{+750}$ & $0.220_{-0.127}^{+0.100}$ & $0.044\pm0.025$ & $0.058_{-0.037}^{+0.036}$ & 2 \\
NGC595 & E--HII & $7390_{-400}^{+390}$ & $8150_{-140}^{+210}$ & $0.320_{-0.166}^{+0.103}$ & $0.031\pm0.015$ & $0.044_{-0.030}^{+0.027}$ & 5 \\
NGC604 & E--HII & $8050_{-200}^{+140}$ & $8640_{-240}^{+260}$ & $0.199_{-0.081}^{+0.072}$ & $0.025\pm0.010$ & $0.030_{-0.016}^{+0.013}$ & 5 \\
NGC6822 & E--HII & $11510_{-260}^{+190}$ & $11340_{-1720}^{+1290}$ & $0.230_{-0.076}^{+0.082}$ & $0.055\pm0.018$ & $0.069_{-0.030}^{+0.027}$ & 10 \\
NGC6822V & E--HII & $11700_{-210}^{+280}$ & $14520_{-2280}^{+1970}$ & $0.210_{-0.133}^{+0.132}$ & $0.051\pm0.037$ & $0.063_{-0.060}^{+0.050}$ & 12 \\
SMC-N66A & E--HII & $12550_{-180}^{+180}$ & $12010_{-290}^{+330}$ & $0.140_{-0.054}^{+0.054}$ & $0.041\pm0.015$ & $0.045_{-0.020}^{+0.019}$ & 3 \\
SMC-N81 & E--HII & $12830_{-250}^{+220}$ & $11850_{-280}^{+220}$ & $0.320_{-0.042}^{+0.036}$ & $0.084\pm0.009$ & $0.116_{-0.018}^{+0.017}$ & 3 \\
SMC-N88A & E--HII & $13820_{-190}^{+250}$ & $13020_{-310}^{+280}$ & $0.170_{-0.028}^{+0.022}$ & $0.060\pm0.010$ & $0.070_{-0.011}^{+0.013}$ & 3 \\
M16 & G--HII & $7600_{-130}^{+140}$ & $8280_{-130}^{+230}$ & $0.400_{-0.108}^{+0.104}$ & $0.042\pm0.011$ & $0.069_{-0.029}^{+0.027}$ & 13 \\
M17 & G--HII & $7950_{-90}^{+100}$ & $8740_{-180}^{+220}$ & $0.250_{-0.082}^{+0.082}$ & $0.029\pm0.009$ & $0.040_{-0.017}^{+0.016}$ & 14 \\
M8 & G--HII & $8050_{-100}^{+110}$ & $8330_{-100}^{+100}$ & $0.360_{-0.063}^{+0.076}$ & $0.043\pm0.008$ & $0.067_{-0.018}^{+0.017}$ & 14 \\
NGC3576 & G--HII & $8450_{-50}^{+40}$ & $8710_{-270}^{+160}$ & $0.301_{-0.032}^{+0.032}$ & $0.041\pm0.004$ & $0.059_{-0.009}^{+0.009}$ & 15 \\
NGC3603 & G--HII & $9020_{-120}^{+130}$ & $11040_{-530}^{+650}$ & $0.301_{-0.112}^{+0.114}$ & $0.046\pm0.019$ & $0.066_{-0.041}^{+0.036}$ & 13 \\
Sh2-100 & G--HII & $8280_{-120}^{+100}$ & $8430_{-300}^{+270}$ & $0.210_{-0.082}^{+0.082}$ & $0.028\pm0.011$ & $0.036_{-0.018}^{+0.016}$ & 16 \\
Sh2-288 & G--HII & $9260_{-620}^{+520}$ & $9390_{-350}^{+340}$ & $0.400_{-0.194}^{+0.188}$ & $0.053\pm0.042$ & $0.089_{-0.109}^{+0.083}$ & 16 \\
Sh2-311 & G--HII & $8960_{-100}^{+90}$ & $9230_{-240}^{+170}$ & $0.250_{-0.141}^{+0.132}$ & $0.035\pm0.031$ & $0.049_{-0.057}^{+0.046}$ & 17 \\
HH 529 II & M42 & $8240_{-80}^{+80}$ & $10150_{-510}^{+570}$ & $0.290_{-0.076}^{+0.073}$ & $0.021\pm0.003$ & $0.056_{-0.006}^{+0.006}$ & 18 \\
HH 529 III & M42 & $8600_{-120}^{+110}$ & $11040_{-970}^{+920}$ & $0.360_{-0.095}^{+0.092}$ & $0.021\pm0.003$ & $0.072_{-0.010}^{+0.010}$ & 18 \\
HH 529 nebula (Blagrave) & M42 & $8536_{-33}^{+35}$ & $10672_{-52}^{+53}$ & $0.060_{-0.032}^{+0.032}$ & $0.009\pm0.004$ & $0.030_{-0.003}^{+0.003}$ & 19 \\
HH 529 shock (Blagrave) & M42 & $8366_{-214}^{+252}$ & $8784_{-729}^{+1184}$ & $0.070_{-0.094}^{+0.094}$ & $0.010\pm0.010$ & $0.040_{-0.013}^{+0.014}$ & 19 \\
M42 Esteban+04 & M42 & $8380_{-60}^{+60}$ & $9920_{-220}^{+290}$ & $0.230_{-0.032}^{+0.032}$ & $0.031\pm0.005$ & $0.041_{-0.008}^{+0.008}$ & 20 \\
M42 Mesa-Delgado+09 & M42 & $8180_{-260}^{+140}$ & $9430_{-320}^{+380}$ & $0.170_{-0.086}^{+0.081}$ & $0.022\pm0.011$ & $0.027_{-0.018}^{+0.014}$ & 21 \\
M42 high-ionization region & M42 & $8442_{-52}^{+49}$ & $9724_{-88}^{+101}$ & $0.254_{-0.024}^{+0.024}$ & $0.034\pm0.003$ & $0.047_{-0.005}^{+0.005}$ & 22 \\
M42 low-ionization region & M42 & $8548_{-57}^{+56}$ & $10150_{-130}^{+140}$ & $0.199_{-0.027}^{+0.027}$ & $0.028\pm0.003$ & $0.036_{-0.006}^{+0.006}$ & 22 \\
M42-1 Mendez-Delgado+21a & M42 & $8420_{-60}^{+80}$ & $9920_{-200}^{+210}$ & $0.149_{-0.036}^{+0.032}$ & $0.022\pm0.005$ & $0.026_{-0.006}^{+0.007}$ & 18 \\
M42-1 Mendez-Delgado+21b & M42 & $8020_{-50}^{+60}$ & $8510_{-110}^{+140}$ & $0.210_{-0.032}^{+0.032}$ & $0.026\pm0.004$ & $0.034_{-0.006}^{+0.006}$ & 23 \\
M42-2 Mendez-Delgado+21a & M42 & $8390_{-50}^{+70}$ & $9860_{-170}^{+230}$ & $0.161_{-0.022}^{+0.022}$ & $0.022\pm0.003$ & $0.027_{-0.004}^{+0.004}$ & 18 \\
M42-2 Mendez-Delgado+21b & M42 & $8120_{-70}^{+50}$ & $8450_{-150}^{+140}$ & $0.369_{-0.045}^{+0.041}$ & $0.044\pm0.005$ & $0.068_{-0.013}^{+0.012}$ & 23 \\
M42-2 Mendez-Delgado+22b & M42 & $8500_{-70}^{+60}$ & $9810_{-200}^{+230}$ & $0.250_{-0.022}^{+0.022}$ & $0.035\pm0.003$ & $0.047_{-0.005}^{+0.005}$ & 24 \\
M42-3 Mendez-Delgado+21a & M42 & $8510_{-60}^{+60}$ & $10090_{-300}^{+230}$ & $0.179_{-0.032}^{+0.036}$ & $0.027\pm0.004$ & $0.033_{-0.007}^{+0.006}$ & 18 \\
M42-3 Mendez-Delgado+22b & M42 & $8410_{-60}^{+50}$ & $9890_{-260}^{+190}$ & $0.250_{-0.022}^{+0.028}$ & $0.033\pm0.003$ & $0.046_{-0.006}^{+0.005}$ & 24 \\
M42-4 Mendez-Delgado+21a & M42 & $8320_{-60}^{+90}$ & $9890_{-210}^{+220}$ & $0.179_{-0.028}^{+0.028}$ & $0.024\pm0.004$ & $0.029_{-0.006}^{+0.006}$ & 18 \\
M42-Bar Delgado-Inglada+16 & M42 & $8470_{-70}^{+90}$ & $9000_{-150}^{+130}$ & $0.190_{-0.063}^{+0.063}$ & $0.026\pm0.008$ & $0.033_{-0.014}^{+0.013}$ & 25 \\
M42-P1 Delgado-Inglada+16 & M42 & $8320_{-60}^{+80}$ & $9500_{-260}^{+250}$ & $0.161_{-0.054}^{+0.063}$ & $0.022\pm0.008$ & $0.028_{-0.011}^{+0.011}$ & 25 \\
\enddata
\tablecomments{The table contains the 59 measurements used in the analysis: 33 extragalactic H II regions (E--HII), eight non-M42 Galactic H II regions (G--HII), and 18 M42 regions or kinematic components. Uncertainties are $1\sigma$. The ADF is $\log(\Opp/\Hp)_\mathrm{RL}-\log(\Opp/\Hp)_\mathrm{CEL}$; its uncertainty is propagated in quadrature from the RL and CEL abundance errors. The analytical/literature column retains the classical Peimbert value quoted or adopted by the source analysis where available; otherwise it is calculated consistently from the published diagnostics. Numerical values integrate a Gaussian temperature distribution using the published line fluxes. Mrk~1259 has no numerical solution for $t^2\leq0.5$.}
\tablerefs{
  (1) \citealt{chen26};
  (2) \citealt{toribio-san-cipriano16};
  (3) \citealt{dominguez-guzman22};
  (4) \citealt{peimbert03};
  (5) \citealt{esteban09};
  (6) \citealt{guseva11};
  (7) \citealt{valerdi19};
  (8) \citealt{pena-guerrero12};
  (9) \citealt{lopez-sanchez07};
  (10) \citealt{esteban14};
  (11) \citealt{esteban20};
  (12) \citealt{peimbert05};
  (13) \citealt{garcia-rojas06};
  (14) \citealt{garcia-rojas07b};
  (15) \citealt{garcia-rojas04};
  (16) \citealt{esteban17};
  (17) \citealt{garcia-rojas05};
  (18) \citealt{mendez-delgado21a}, HH 529;
  (19) \citealt{blagrave06};
  (20) \citealt{esteban04};
  (21) \citealt{mesa-delgado09};
  (22) \citealt{chen26orion};
  (23) \citealt{mendez-delgado21b}, HH 204;
  (24) \citealt{mendez-delgado22}, HH 514;
  (25) \citealt{delgado-inglada16}.
}
\end{deluxetable*}
\end{longrotatetable}